\documentclass[sn-mathphys,Numbered]{sn-jnl}

\usepackage{graphicx}%
\usepackage{multirow}%
\usepackage{amsmath,amssymb,amsfonts}%
\usepackage{amsthm}%
\usepackage{mathrsfs}%
\usepackage[title]{appendix}%
\usepackage{xcolor}%
\usepackage{textcomp}%
\usepackage{manyfoot}%
\usepackage{booktabs}%
\usepackage{algorithm}%
\usepackage{algorithmicx}%
\usepackage{algpseudocode}%
\usepackage{listings}%

\newcommand\splot[3][0.26]{\parbox{#1\textwidth}{\centering
    \includegraphics[width=#1\textwidth]{#2}\par
    \strut(#3)}}
\DeclareMathOperator\erfc{erfc}
\DeclareMathOperator\erf{erf}
\newcommand{\diff}{\mathrm{d}}

\begin{document}
\title{Scale dependence of distributions of hotspots}

\author*[1]{\fnm{Michael}
  \sur{Wilkinson}}\email{m.wilkinson@open.ac.uk}
\author[2,3]{\fnm{Boris} \sur{Veytsman}}\email{bveytsman@chanzuckerberg.com}

\affil*[1]{\orgdiv{School of Mathematics and Statistics}, \orgname{The
    Open University}, \orgaddress{\street{Walton Hall}, \city{Milton
      Keynes}, \postcode{MK7 6AA}, \country{UK}}}
\affil[2]{\orgname{Chan Zuckerberg Initiative},
  \orgaddress{\street{1180 Main St}, \city{Redwood City},
    \postcode{94063}, \state{CA}, \country{USA}}}
\affil[3]{\orgdiv{School of Systems Biology}, \orgname{George Mason
    University}, \orgaddress{\street{4400 University Dr},
    \city{Fairfax}, \postcode{22030}, \state{VA}, \country{USA}}}

\abstract{%
  We consider a random field $\phi(\mathbf{r})$ in $d$ dimensions
  which is largely concentrated around small `hotspots', with 
  `weights', $w_i$. These weights may have a very broad distribution, 
  such that their mean does not exist, or else is not a useful estimate. In such  
  cases, the median $\overline W$ of the total weight
  $W$ in a region of size $R$ is an informative characterisation of the weights. 
  We define the function $F$ by
  $\ln \overline W=F(\ln R)$. If $F'(x)>d$, the distribution of hotspots is
  dominated by the largest weights.  In the case where $F'(x)-d$
  approaches a constant positive value when $R\to \infty$, the hotspots
  distribution has a type of scale-invariance which is different from
  that of fractal sets, and which we term \emph{ultradimensional}. The
  form of the function $F(x)$ is determined for a model of diffusion
  in a random potential.
}

\maketitle

\section{Introduction}
\label{sec: 1}

In many cases two-dimensional scalar fields are largely supported on
small areas, `hotspots'. Examples can include the distribution of
human populations, which are concentrated in urban settlements, the
distribution of debris on the ocean, which can be concentrated in
regions where cool or saline water is subducted, and deposits of
mineral ores, which can be concentrated at the points where dissolved
material is deposited from evaporating water. Another example is 
images of star fields, where the stars appear as points. Subjectively,
images of the distribution of hotspots can appear to bear a familial
resemblance. This paper addresses the question of how these
distributions can be characterised, and whether they have
scale-invariant features.

The fields that we consider can be modelled by random processes. We
consider random, non-negative scalar fields in a two-dimensional
space, denoted by $\phi(\mathbf{r})$, with statistics which are
homogeneous (translationally invariant) and isotropic (rotationally
invariant).  Extensions to higher dimensions will be obvious.

In the cases where the field is highly concentrated in the vicinity of
isolated points, we can consider the following simple model. We take a
uniform, independent random scatter of points on the plane,
$\mathbf{r}_i$, with density $\rho$. Each point is assigned a
random weight $w_i$, drawn independently from a distribution with 
probability density function (PDF)
$p(w)$.  The weights $w_i$ represent the integral of
$\phi(\mathbf{r})$ in the neighbourhood surrounding one of the points
upon which it is concentrated.

The primary interest will be in the cases where $p(w)$ has a very
broad distribution.  Accordingly, we introduce a \emph{power-law
  model}, such that for large $w$, the PDF is $p(w)\sim
w^{-\gamma}$. In the calculations below we shall use the following
specific distribution as an example:
\begin{equation}
  \label{eq:p(w)}
  p(w)=
  \begin{cases}
    (\gamma-1)w^{-\gamma},&\quad w\ge 1\\
    0,&\quad w<1
  \end{cases}
\end{equation}
with $1<\gamma <2$, so that the distribution is normalisable, but its
mean is undefined.

We shall also need to consider the cumulative distribution: if
$P(w_0)$ is the probability that $w>w_0$, then (\ref{eq:p(w)}) implies
that $P(w)= w^{-(\gamma -1)}$ for $w>1$. It will be argued that this
is a foundational model for the distribution of hotspots, and that
distributions obtained from more general models can be approximated
using this model, with a suitable choice of the parameter $\gamma$.

\begin{figure}
  \centering
  \splot{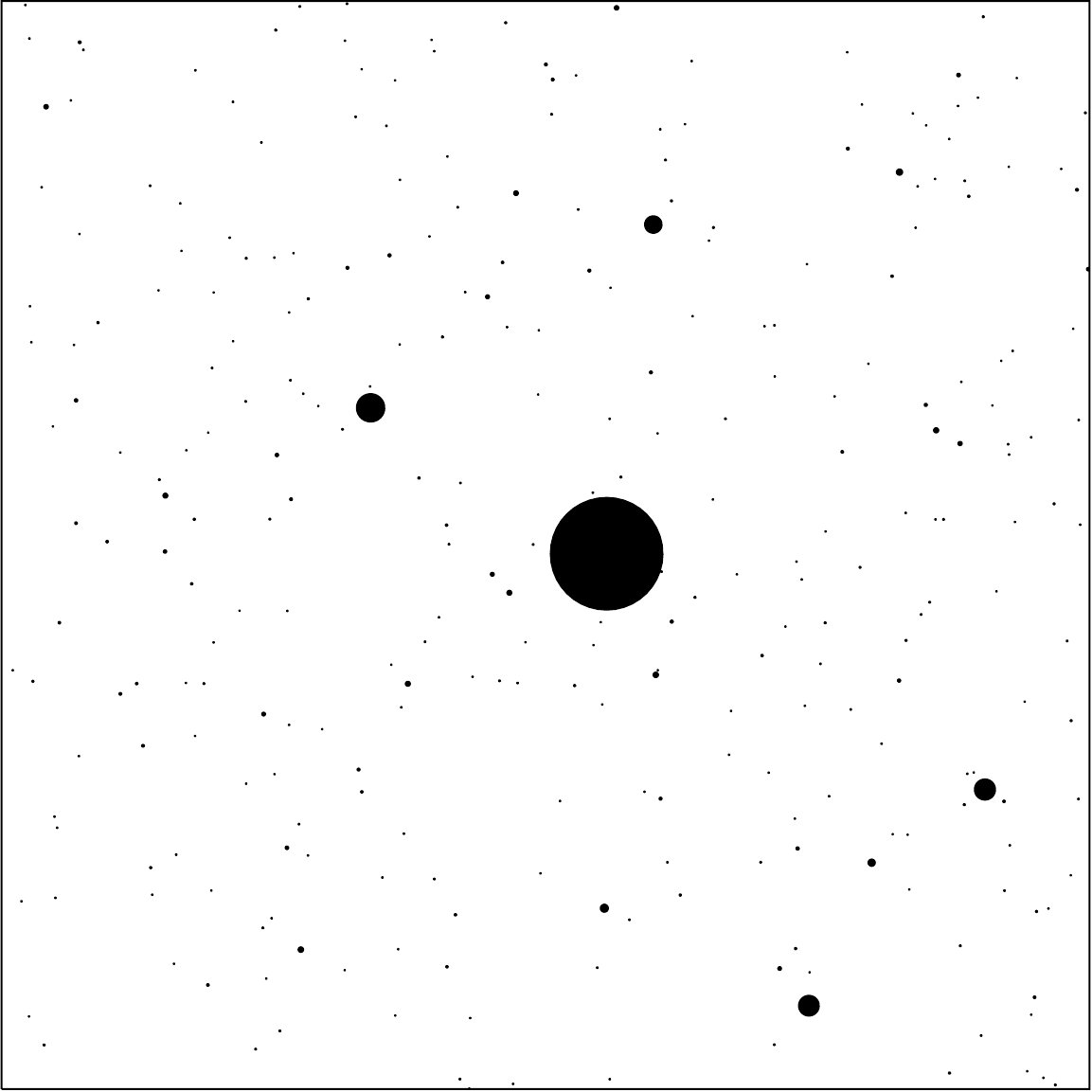}{a}
  \splot{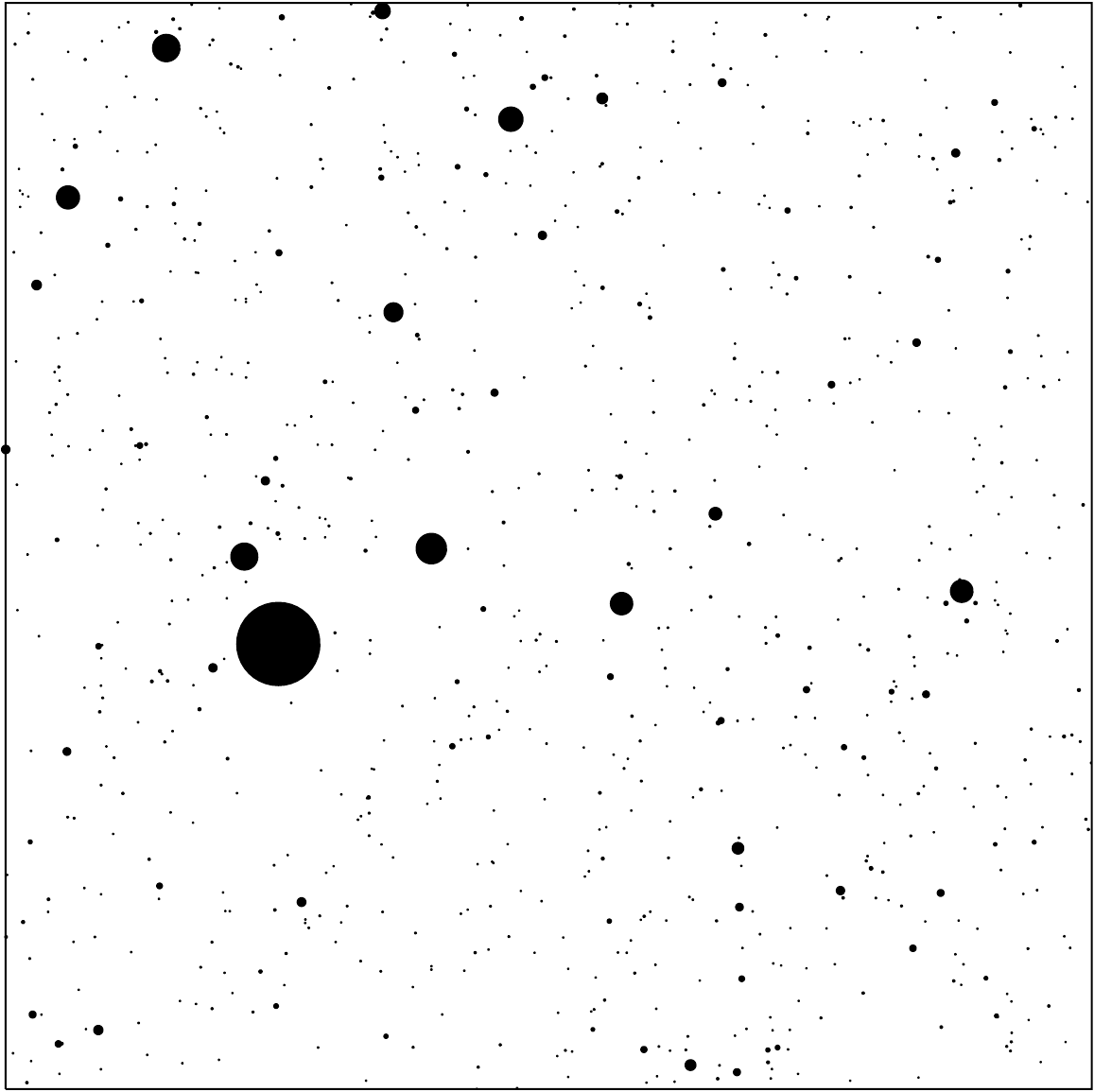}{b}
  \splot{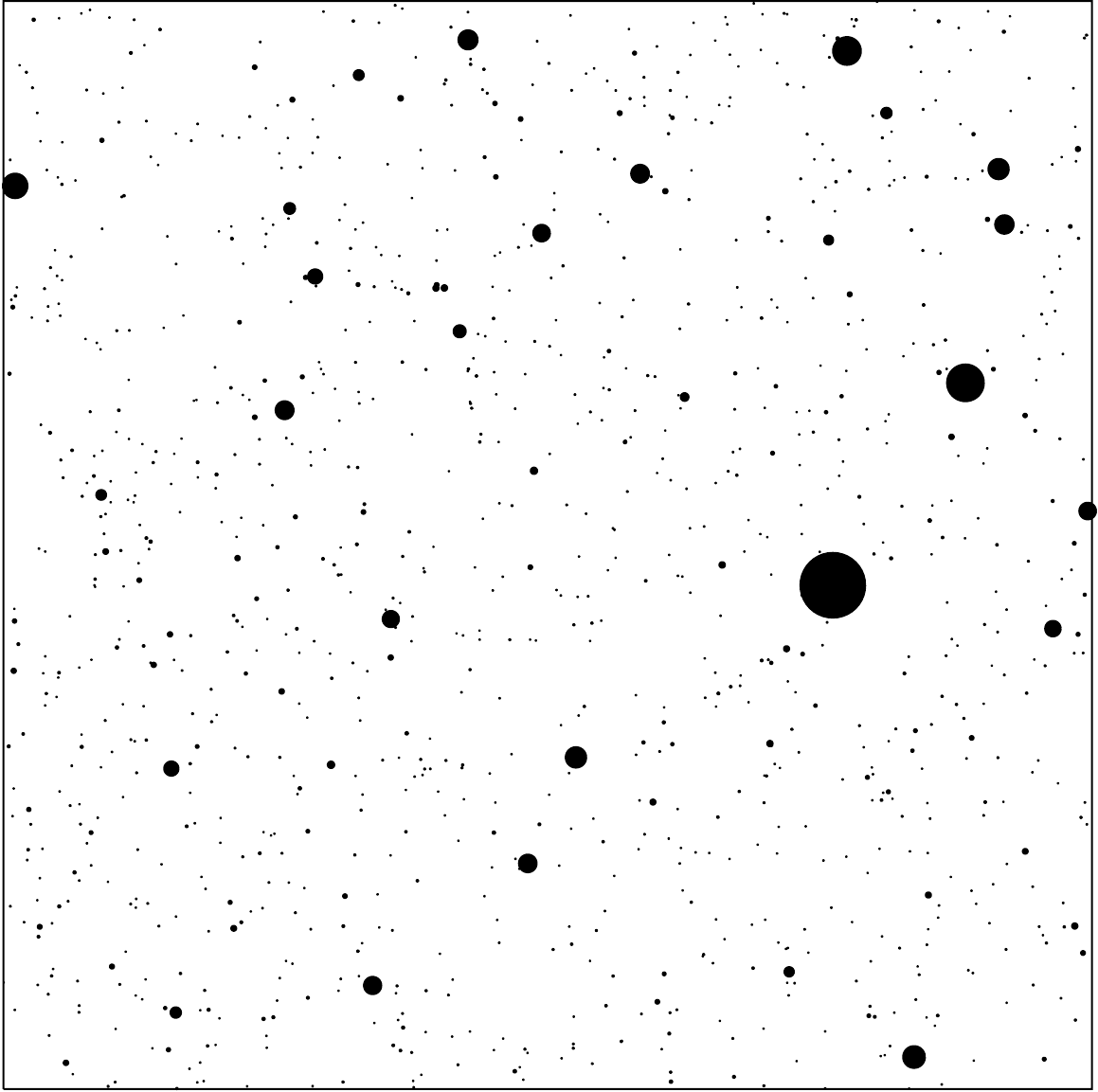}{c}
  \splot{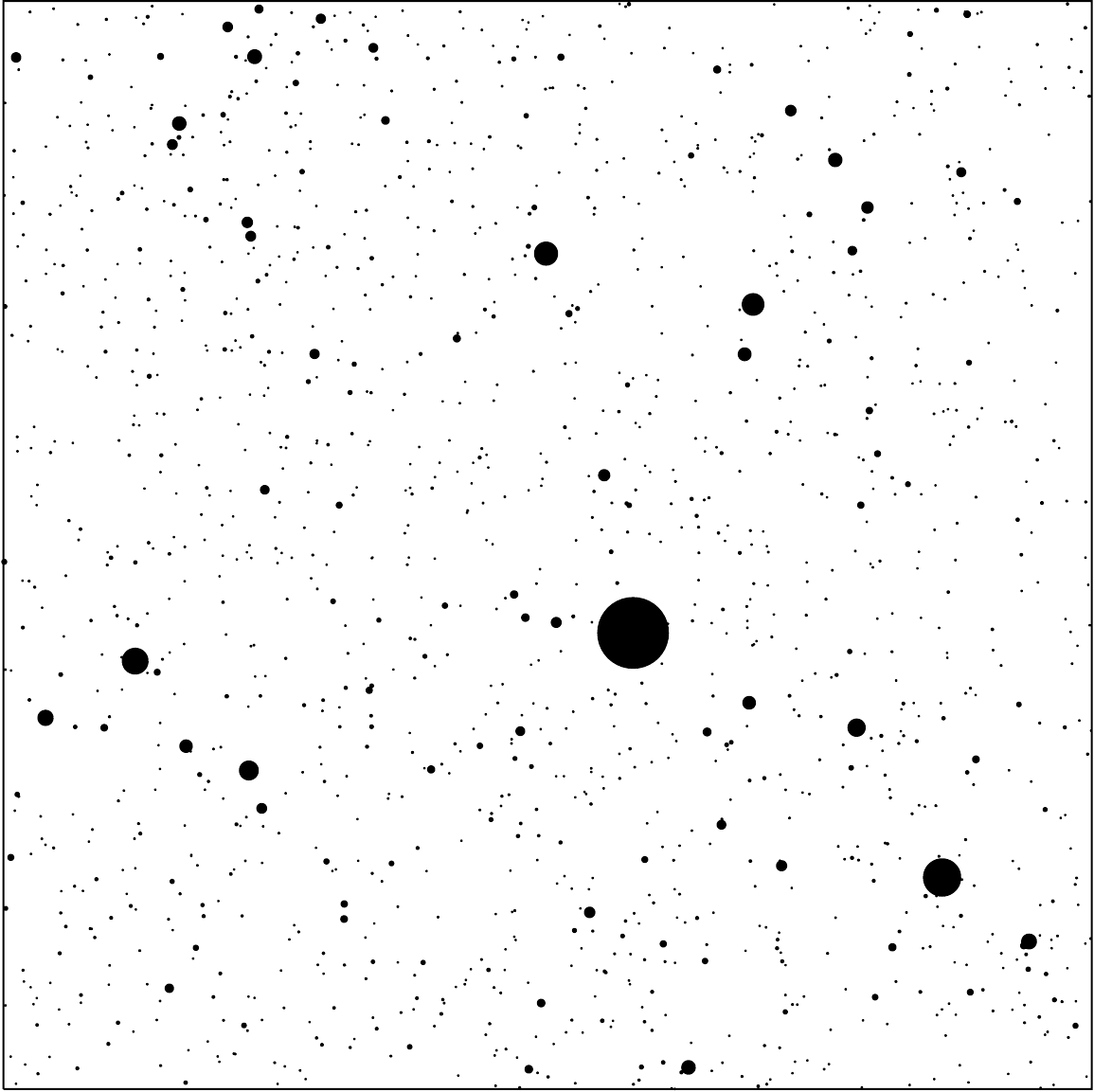}{d}
  \splot{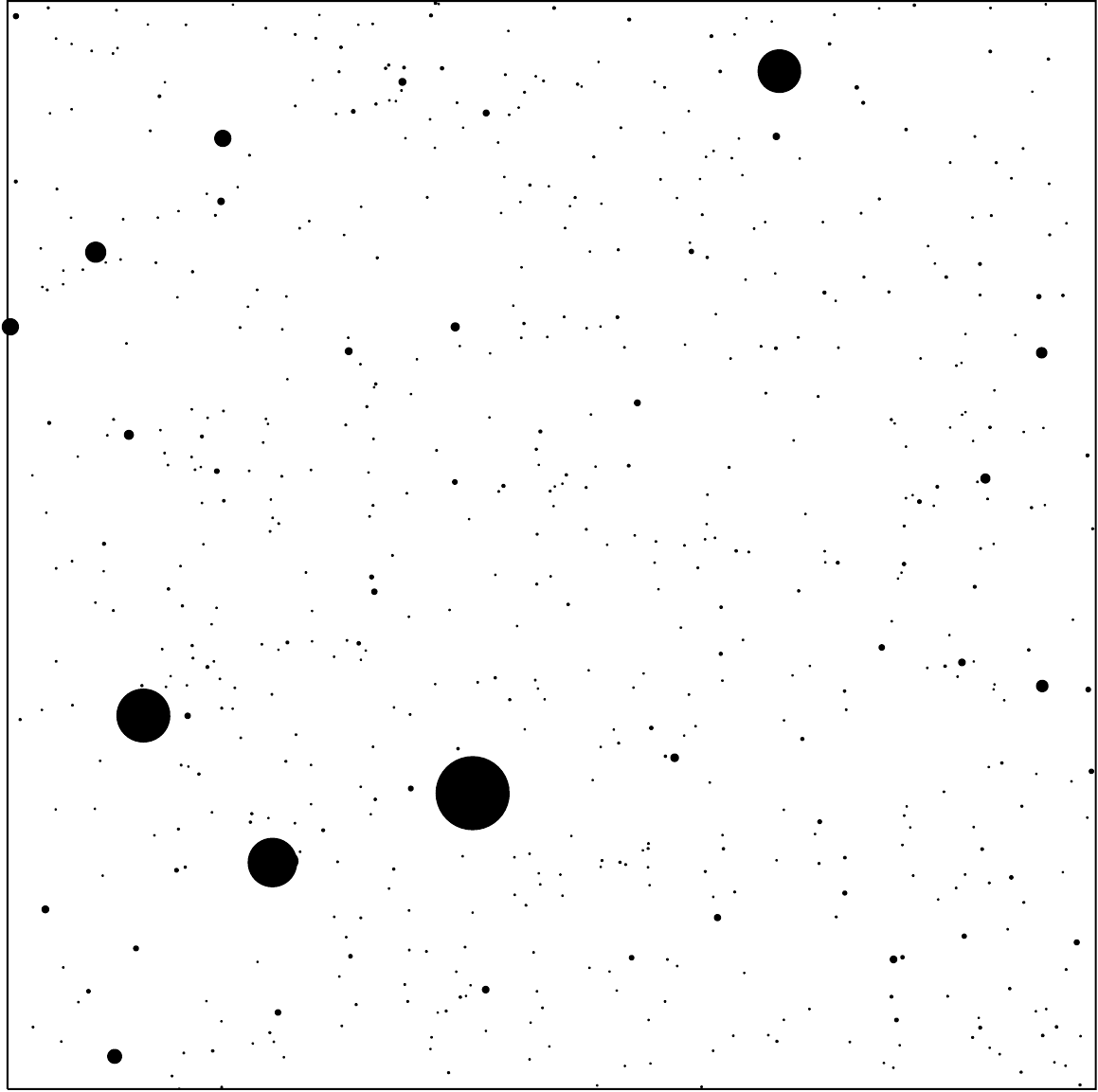}{e}
  \splot{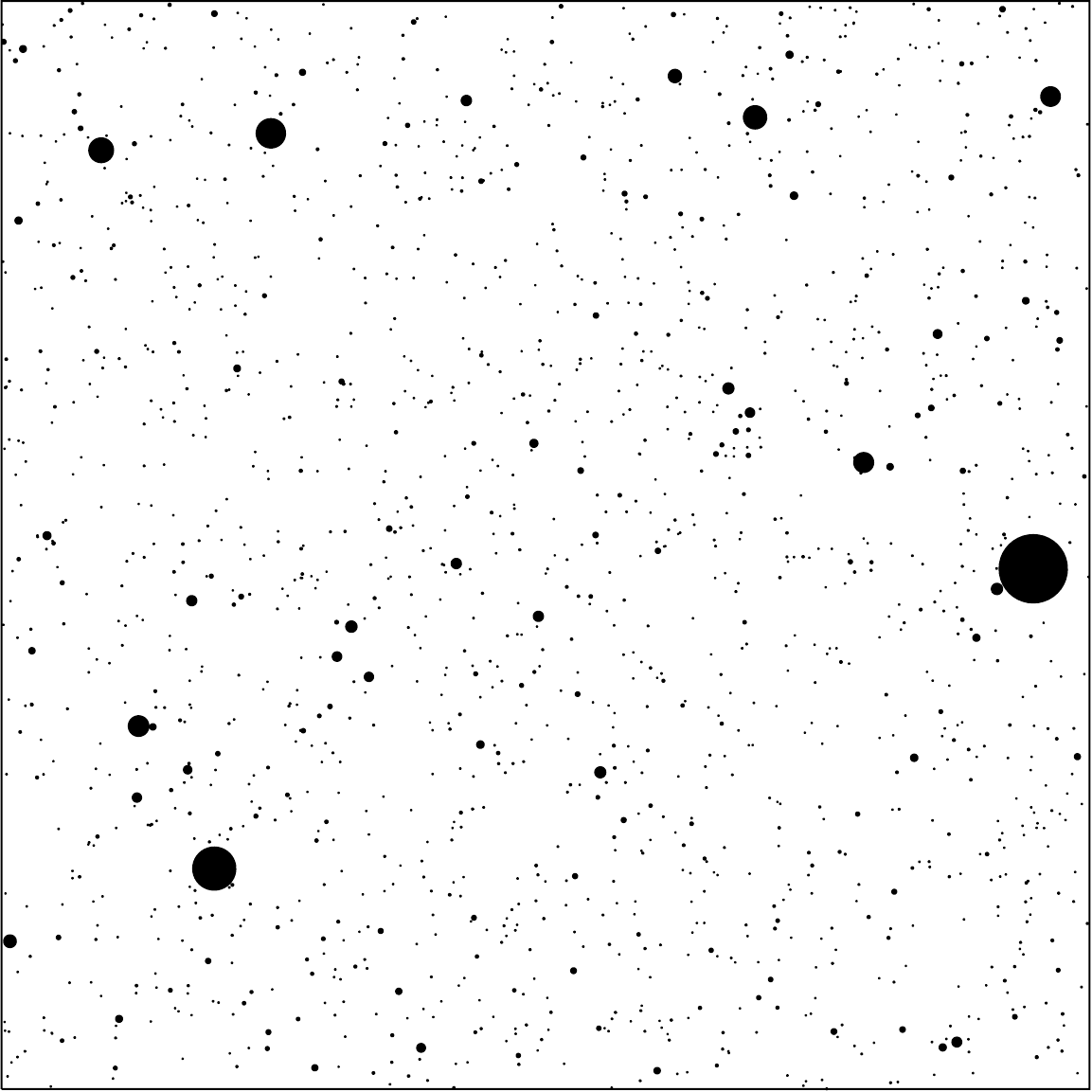}{f}
  \splot{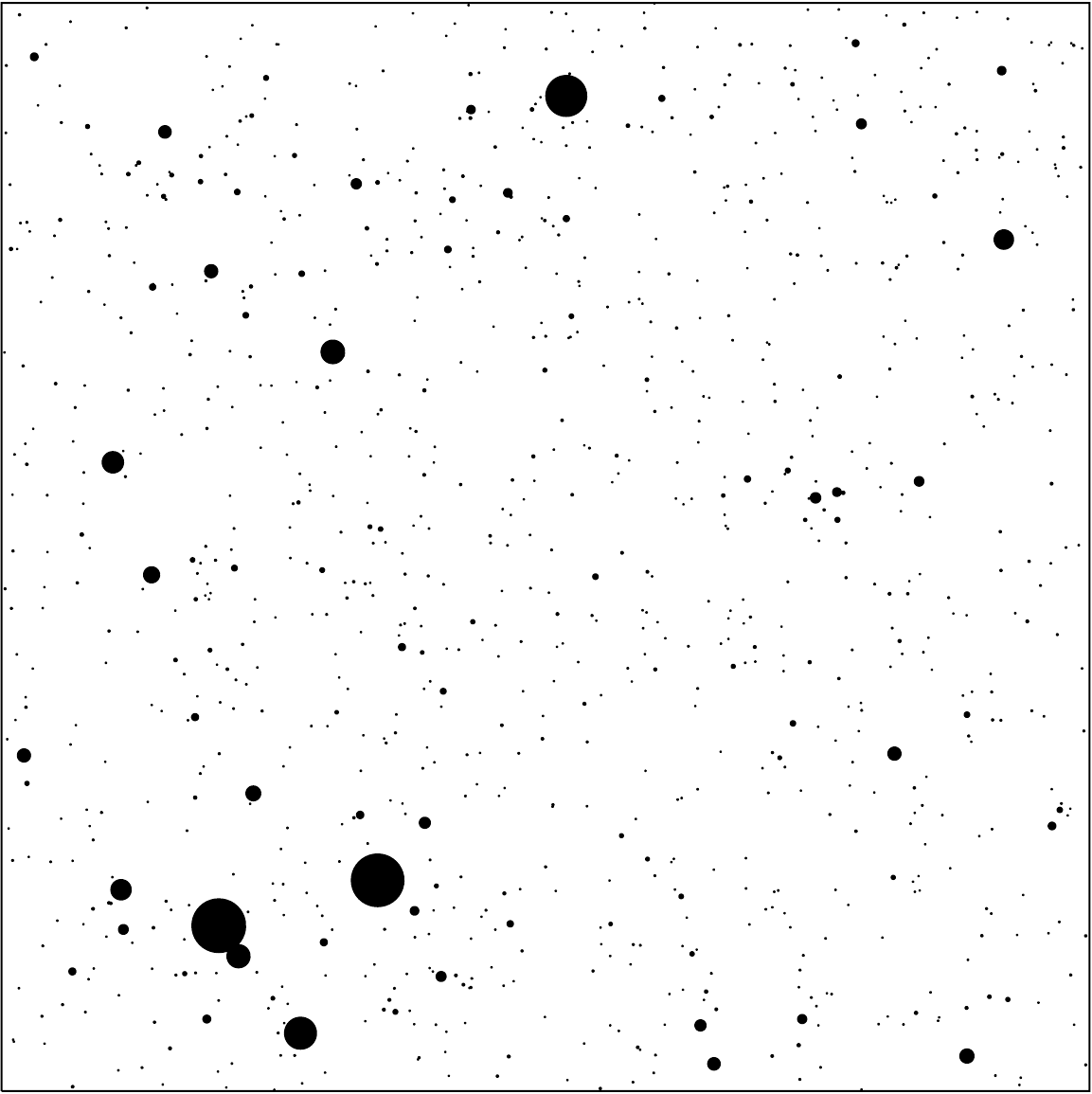}{g}
  \splot{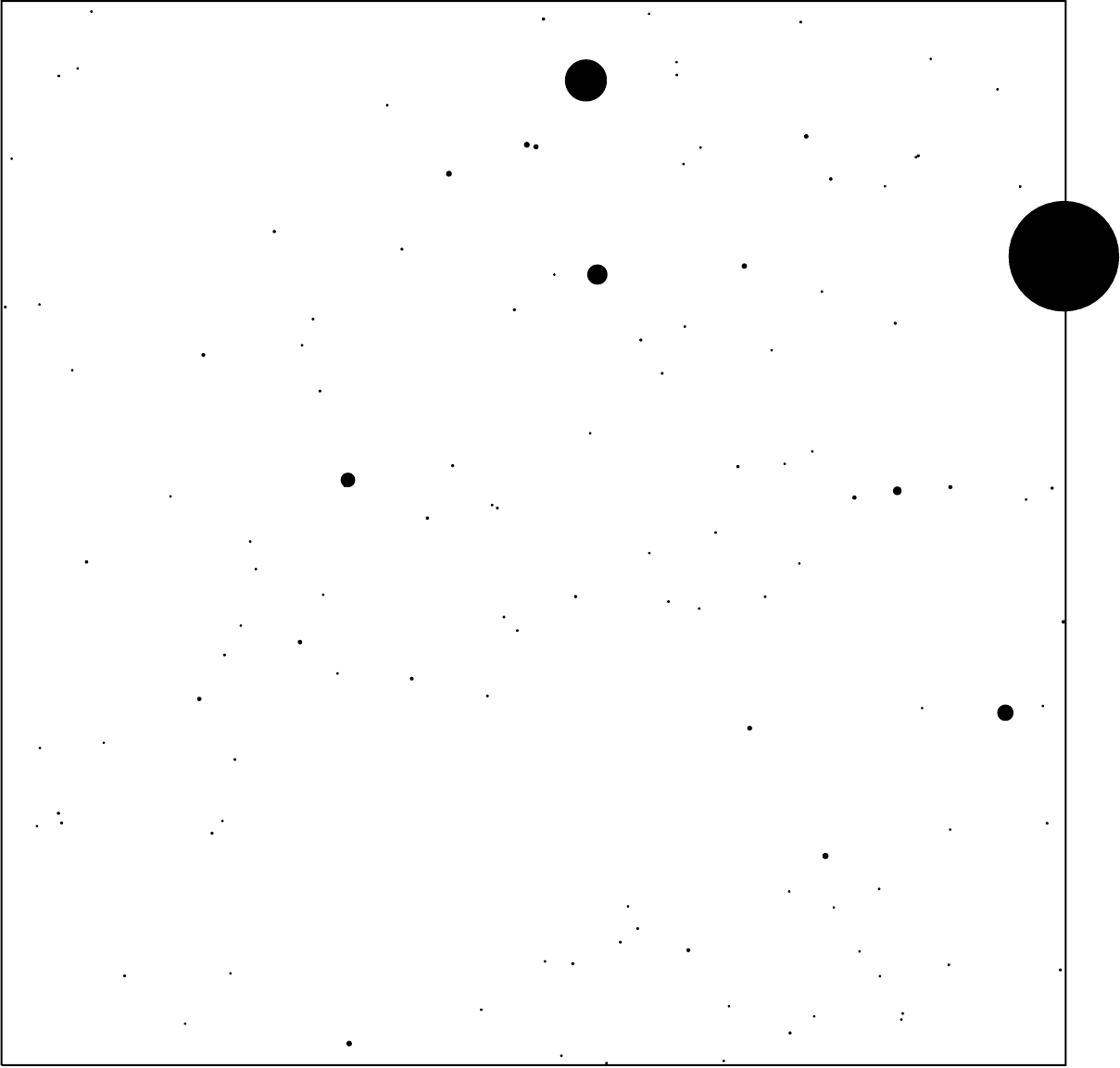}{h}
  \splot{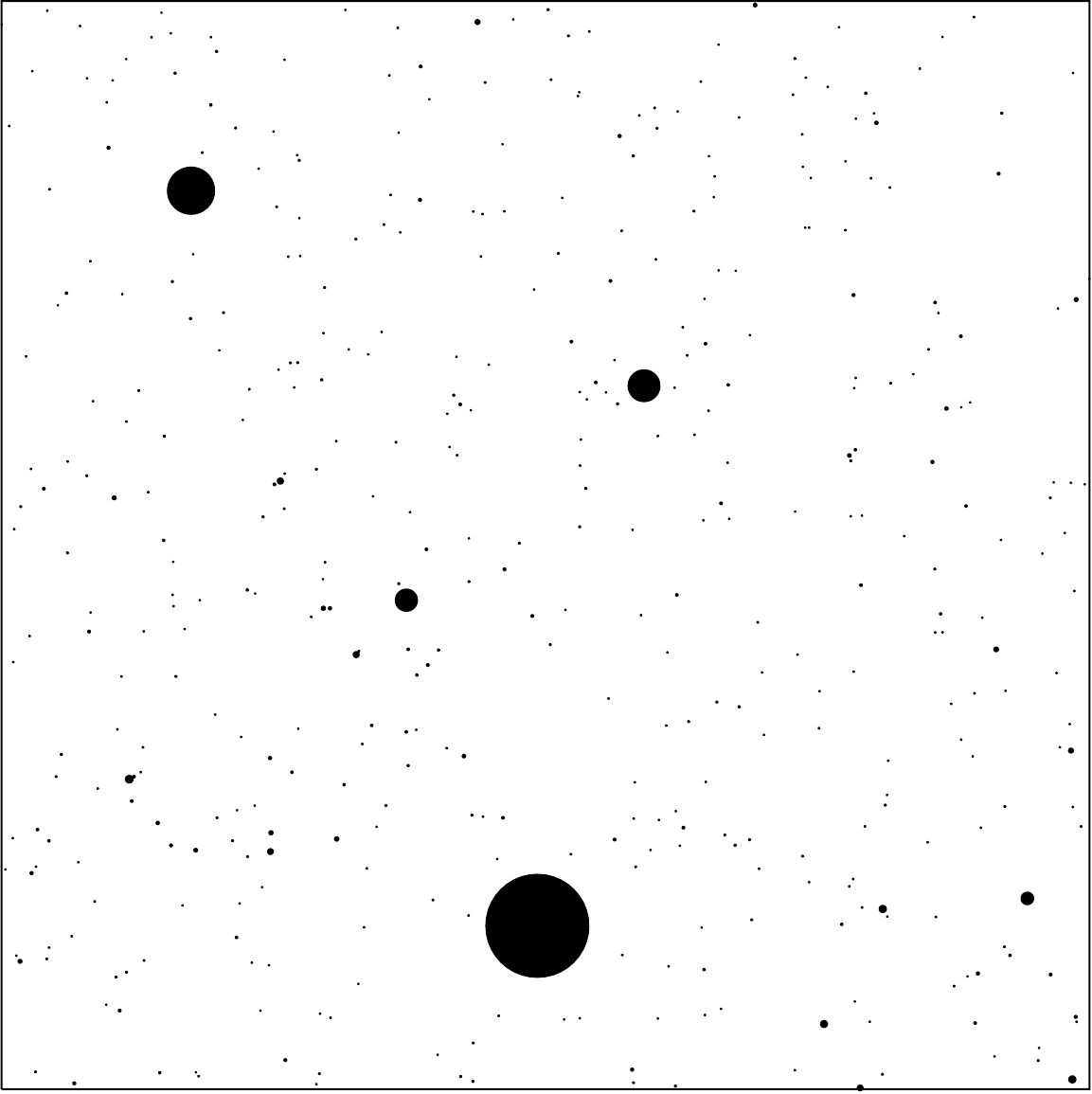}{i}
  \splot{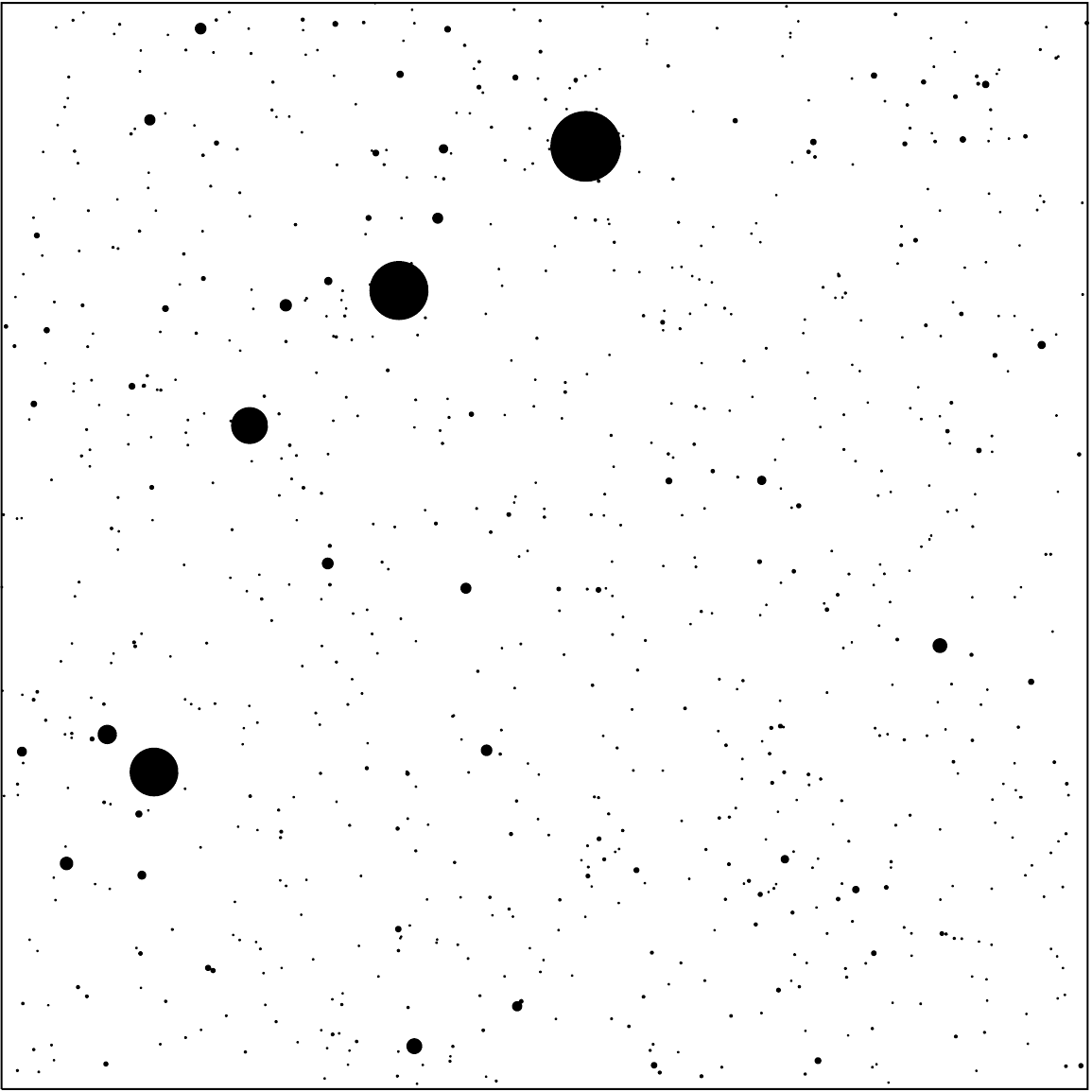}{j}
  \splot{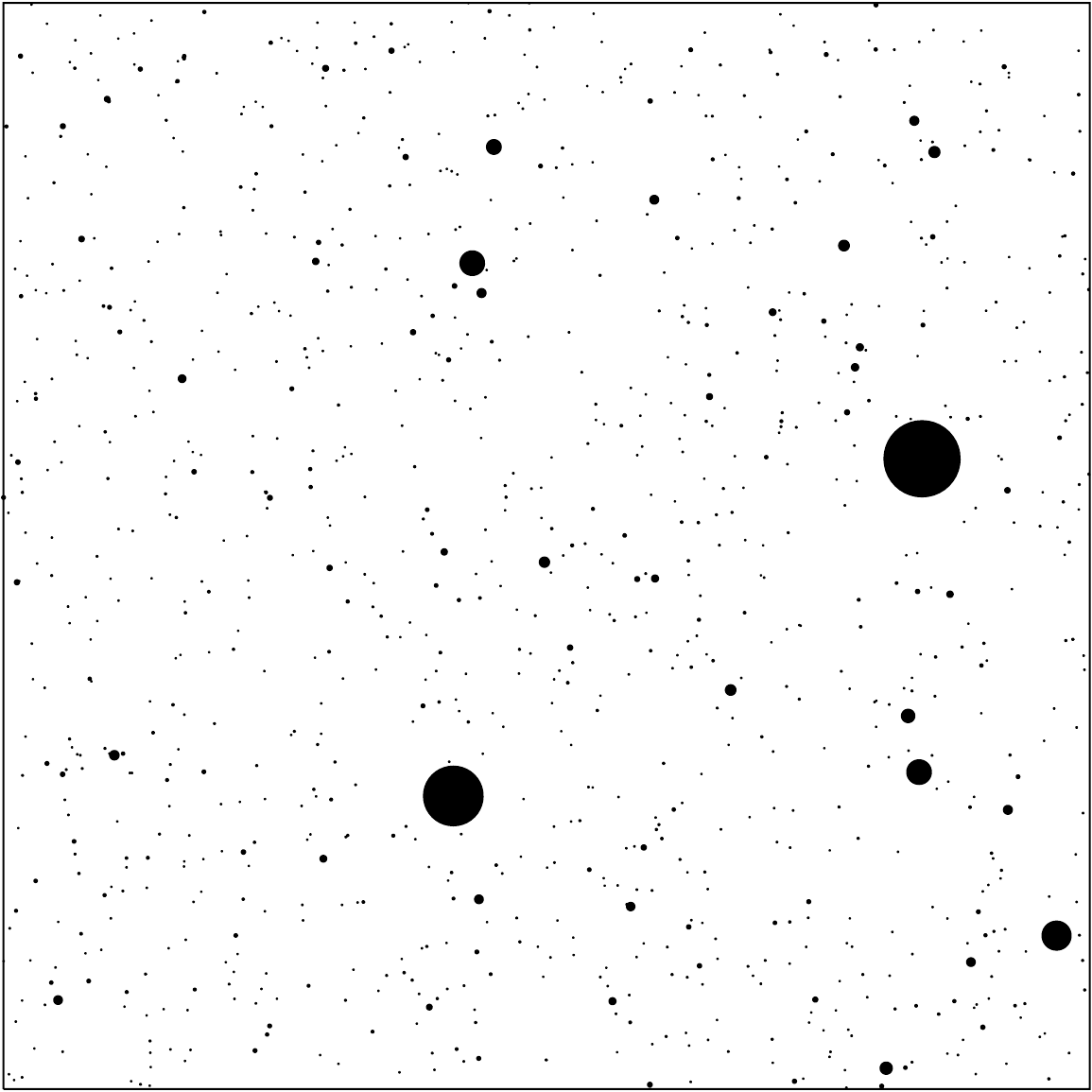}{k}
  \splot{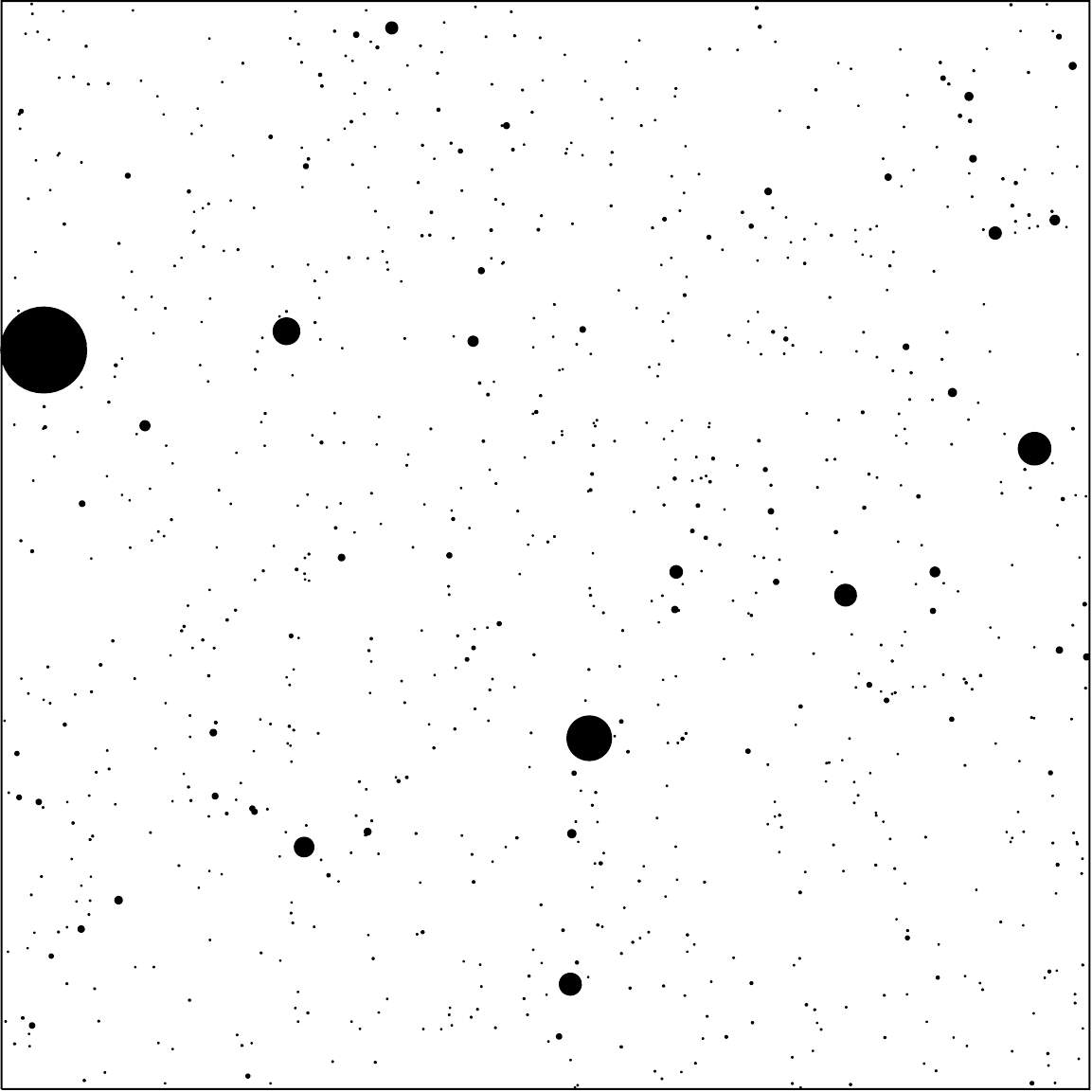}{l}
  \caption{Illustration of scale-independence of the `hotspots' model.
    At the position of each hotspot there is a filled circle with
    area proportional to its weight.  The total area of the circles in
    each image is normalised to be $1\%$ of the total area of the
    image.  The images use the probability distribution
    (\ref{eq:p(w)}) with $\gamma=5/3$, and the scale factors are
    $R=10000$, $R=2000$, $R=400$, $R=80$, (with three cases of each
    scale factor). The different images cannot be associated with the
    different values of $R$ by any statistical test, reflecting the
    scale-invariance property.  The scale factors are: $R=10000$,
    panels (a), (g), (k), $R=2000$, panels (b), (f), (h), $R=400$,
    panels (d), (e), (l), $R=80$, panels (c), (i), (j).  }
\label{fig:hotspots}
\end{figure}

Figure \ref{fig:hotspots} is an illustration of $12$ different
realisations of this model for hotspot distributions, with
$\gamma=5/3$, plotted on four different lengthscales. It will be
argued that the statistics of these images has a scale-invariance
property, in that it is impossible to identify the scale factors of
the panels.  Non-trivial scale invariance is usually associated with
fractal~\cite{Man83,Fal90} (or more generally, multifractal
\cite{Hal+86,Sal+17}) properties, which can usually be characterised
by saying that the set is, in some sense, self-similar under a change
of scale. The images in figure \ref{fig:hotspots} are so diverse that
would require a large number of realisations to demonstrate that they
are drawn from the same ensemble. We shall argue below that there is a
simple quantitative distinction between the scale invariance of figure
\ref{fig:hotspots} and that of fractal sets.

There are two complementary aspects to characterising these
sets. Firstly, we might wish to know how the total weight of the
hotspots increases with the size of the region.  Consider the set
$\mathcal{W}_R$ of approximately $\rho R^2$ values of $w_i$ for which
$\mathbf{r}_i$ lies inside a square of side $R$.  The set has the
total weight
\begin{equation}
\label{eq:W}
W=\sum_{w_i\in\mathcal{W}_R}w_i.
\end{equation}
If the weights had a compact distribution, we would estimate the mean
value of $W$ as $\langle W\rangle =\rho R^2\langle w\rangle$, but for
the distribution~(\ref{eq:p(w)}), $\langle w\rangle$ is infinite ($\langle X\rangle$ denotes 
the expectation value of $X$ throughout). A
more promising approach is to estimate the median value $\overline W$. This
is considered in Section~\ref{sec:simple_model} below. We anticipate
that $\overline W$ will increase very rapidly as a function of the scale
length $R$. Accordingly, we use a logarithmic scale.  We can
characterise a given hotspot distribution by means of a function $F$:
\begin{equation}
\label{eq:ln_bar_w}
\ln(\overline W)=F(\ln R).
\end{equation}
For the simple model described above, we show that $\overline W$ has a
power law dependence upon $R$, so $F(x)$ is a linear function. The
exponent of this power law can be thought of as a type of
\emph{dimension} of the set of hotspots, and even if $F$ is not a
linear function we can define an effective dimension $D_\text{eff}$ at
the length scale $R$ as the derivative
 \begin{equation}
 \label{eq:Deff}
D_\text{eff}=\frac{\diff F(\ln R)}{\diff (\ln R)}.
 \end{equation}
For our simplified model it will be shown that, for points distributed 
randomly in $d$ dimensions with the weight distribution (\ref{eq:p(w)}),
\begin{equation}
\label{eq:Deff1}
D_\text{eff}=\frac{d}{\gamma -1}.
\end{equation}
Note that, because $2>\gamma>1$, this effective dimension is higher
than the dimension of the embedding space. This indicates that the
effective dimension is different from a fractal dimension. We describe
this scale invariance as \emph{ultradimensional}.

A second aspect of describing the hotspot distribution is to consider
the relative sizes of the largest values of $w_i$ in the set
$\mathcal{W}_R$. We can transform this to a filtered and normalised
set, $\widetilde{\mathcal{W}}_R$, as follows.  We scale the hotspot
positions by dividing by $R$, and plot $\mathbf{r}_i/R$ inside a unit
square.  We eliminate the values of $w_i$ below a chosen threshold,
for example, those $w_i$ that are less than $\epsilon W$, where
$\epsilon$ is a given small positive number. We can also `normalise'
these sets by dividing every remaining $w_i$ by $W$.  These normalised
and filtered sets are a natural representation of many types of
point-set data.  An example is a geographical map showing settlements
using symbols with the sizes relative to of the largest settlements in
the mapped region, where settlements below a certain size are not
shown in order to eliminate clutter. Another example is a photograph
of the night sky with the exposure adjusted so that the image
saturation is normalised, and stars below a certain intensity are not
registered at all. The filtered and normalised sets can be
characterised by considering the relative sizes of the largest values
of $w_i$. To this end, we can sort the weights, $w_i$, into a
decreasing sequence $\{ \overrightarrow w_i\}$, and consider the
proportion of the total mass which is contained in the first $k$
elements of this set
\begin{equation}
\label{eq:f_k}
f_k=\frac{\sum_{j=1}^k  \overrightarrow w_j}
{\sum_{j=1}^{\widetilde N}  \overrightarrow w_j},
\end{equation}
where $\widetilde N$ is the number of elements in the filtered set.  We
can consider the average of $f_k$ over different regions of the data,
and in some cases we can also average over multiple realisations of
the distribution.  For the model defined by equation (\ref{eq:p(w)}),
this leads to a family of functions of $\gamma$:
\begin{equation}
\label{eq:widetilde_f_k}
\widetilde f_k(\gamma)=\langle f_k \rangle.
\end{equation}
We shall make a hypothesis that, for a general model, the set of
values of $f_k$ at length scale $R$ is representative of the
model~(\ref{eq:p(w)}), with an effective value of $\gamma$ 
given by rearrangement of~(\ref{eq:Deff1}):
\begin{equation}
\label{eq:gamma_eff}
\gamma_\text{eff}=1+\frac{d}{F'(\ln\,R)}.
\end{equation}

If the derivative of the function $F(x)$ defined by
equation~(\ref{eq:ln_bar_w}) approaches a constant as $x\to\infty$,
this is indicative of the sets $\widetilde{\mathcal{W}}_R$ having
scale-invariant properties, such that the statistics of
$\widetilde{\mathcal{W}}_R$ and $\widetilde{\mathcal{W}}_{\lambda R}$ are
indistinguishable, for a wide range of values of the positive number
$\lambda$. This idea can be expressed by saying that the realisation
of $\widetilde{\mathcal{W}}_R$ are drawn from an ensemble which is
independent of $R$, depending only upon $\gamma_\text{eff}$.  In the
case of the power law model, this scale invariance is manifest.  This
self-similarity could be trivial, or it could indicate that the
hotspot distribution has fractal properties, or something
different. It will be argued that it is the latter possibility which
is realised.  Figure \ref{fig:hotspots} is an example of sets
$\widetilde{\mathcal{W}}_R$ generated by this model.  displayed on four
different length scales (we used $\rho=1$, $\gamma=5/3$, and $R=80$,
$400$, $2000$, $10000$. We show $12$ sample sets, with three at each
of the different scale factors. The twelve images look so different
from each other that it is not evident that they are drawn from the
same ensemble. The different values of the scale factor $R$ were
randomly assigned (the key is in the figure caption), and its value
cannot be determined by inspection of an individual realisation.
Despite the fact that the length scales vary by a factor of $125$, it
is not possible to distinguish which of these images corresponds to
which value of $R$.

The results in section \ref{sec:simple_model} will quantify the
non-trivial scale-invariance of the power-law model under a change of
the magnification of the image.  We discuss the statistics of $W$ for
this simple model, leading to the relation (\ref{eq:Deff1}) between
the exponent $\gamma$ and the dimension $D_\text{eff}$.

In section \ref{sec:diffusion} we discuss a physical example of a
hotspot distribution, namely the probability density for a particle
diffusing in a two-dimensional gaussian random potential, $V(x,y)$.
The equilibrium probability density is proportional to
$\exp[-V(x,y)/\mathcal{D}]$ where $\mathcal{D}$ is the diffusion
coefficient. In the limit where the diffusion coefficient approaches
zero, this density is concentrated at `hotspots' which are minima of
the potential function, with the weights
$w\approx \exp[-V_\text{min}/\mathcal{D}]$, where $V_\text{min}$ is
the value of the local minimum of the potential at the hotspot. We
identity the functions $F(\ln R)$ and $\gamma_\text{eff}(R)$ for this
model, including their dependence upon the diffusion coefficient
$\mathcal{D}$. Section \ref{sec:conclusion} is a brief conclusion.

\section{Statistics of a simple model}
\label{sec:simple_model}

Consider how the statistics of the total weight $W$ depends upon $R$
for the \emph{power-law model}, with weight distribution (\ref{eq:p(w)}).
The mean value of $w$ is undefined, so calculating the expectation
value $\langle W\rangle$ is not a good approach. The mean value is
dominated by rare realisations where one or more of the $w_i$ takes a
very large value. Estimating the median of $W$, which will be denoted
by $\overline W$, appears to be more promising. If $\hat w$ is the largest
of the $N\sim \rho R^2$ samples of $w$ in the square, then we might
hypothesise that $\overline W$ is approximated by $\overline {\hat w}$, that is
of the median of the largest value in the sample.  Here it will be
argued that this multiplier $\overline W/\overline{\hat w}$ is independent of
both $R$ and $\epsilon$.

It is easy to calculate $w^\ast\equiv \overline{\hat w}$. The probability
that none of the $N$ independent values of $w$ exceeds $\hat w$ is
$[1-P(\hat w)]^N$, so that $w^\ast=\overline{\hat w}$ satisfies
$[1-P(w^\ast)]^N=1/2$. This gives
\begin{equation}
\label{eq:w_ast}
w^\ast=\left(\frac{N}{\ln2}\right)^{1/(\gamma-1)}.
\end{equation}

Next we estimate the number of points $\widetilde N$ in the filtered set,
and the value of $\overline W$.  The number of values of $w_i$ in the range
from $w^\ast \equiv \overline {\hat w}$ (upper limit) to $\epsilon w^\ast$
(lower limit) is
\begin{equation}
  \label{eq:widetildeN}
  \begin{aligned}
    \widetilde N&\sim N\int_{\epsilon w^\ast}^{w^\ast}\diff w\, p(w)\\
    &= N\left[w^{1-\gamma}\right]_{\epsilon w^\ast}^{w^\ast}\\
    &\sim N (w^\ast)^{-(\gamma-1)}\epsilon^{-(\gamma-1)}\\
    &\sim \ln2\, \epsilon^{-(\gamma-1)}
  \end{aligned}
\end{equation}
so that the number of points $\widetilde N$ in the filtered set is
independent of $R$, although it does depend upon $\epsilon$.

The median of the sum $W$ of a large number of values of $w_i$ is
estimated by noting that $W=\hat w+\widetilde W$, where $\hat w$ is the
largest of the $w_i$, and $\widetilde W$ is the sum excluding the largest
of the $w_i$. The value of $\widetilde W$ will be approximated by its mean
value, which depends upon $\hat w$. Writing $\hat w=aw^\ast$, and
taking the leading order as $N\to \infty$, $\epsilon\to 0$
\begin{equation}
  \label{eq:averW}
  \begin{aligned}
    \langle \widetilde W\rangle&\sim (N-1)\int_{\epsilon w^\ast}^{\hat
      w}\diff w\, w p(w)\\ 
    &\sim (N-1)
    \left[\frac{\gamma-1}{2-\gamma}w^{2-\gamma}\right]_{\epsilon \hat
      w}^{\hat w} \\
    &\sim  \frac{\gamma-1}{2-\gamma}(N-1)\hat w^{2-\gamma} \\
    &\sim  \frac{(\gamma-1)\ln 2}{2-\gamma}a^{2-\gamma }w^\ast.
  \end{aligned}
\end{equation}

This gives the following estimate for $W$, in terms of $a=\hat w/w^\ast$:
\begin{equation}
\label{eq:W_estimate}
W\approx w^\ast\left[a+a^{2-\gamma}\frac{(\gamma-1)\ln 2}{2-\gamma}\right].
\end{equation}

The value of $W$ depends upon a random quantity, $a$. The median value
of $W(a)$ is $\overline W=W(\overline a)$. And because we define
$w^\ast =\overline {\hat w}=w^\ast \overline a$, we have
$\overline a=1$. This gives the following estimate for $\overline W$:
\begin{equation}
\label{eq:bar_W_estimate}
\overline W\approx \left(1+\frac{(\gamma -1)\ln 2}{2-\gamma}\right)\left(\frac{N}{\ln\, 2}\right)^{1/(\gamma-1)}.
\end{equation}

This indicates that $\overline W$ exceeds the median of the largest
term by a factor which is independent of both $\epsilon$ and $N$ (and
which is therefore therefore independent of $R$). The independence of
$\overline W/w^\ast$ upon $R$ indicates that the filtered images are
scale-invariant. The fact that this ratio does not depend upon
$\epsilon$ reflects the fact that the images are dominated by the
largest values of $w_i$. Equation~(\ref{eq:bar_W_estimate}) implies
that the number of $w_i$, including the largest one, that make a
significant contribution to $\overline W$ is
$1+\frac{\gamma -1}{2-\gamma}\ln 2$.  When $\gamma\to 1$, there is
likely to be only one $w_i$ that dominates the filtered image.  This
is in accord with the \emph{large jump principle}, discussed in
\cite{Vez+19}.

\begin{figure}
  \centering
  \includegraphics{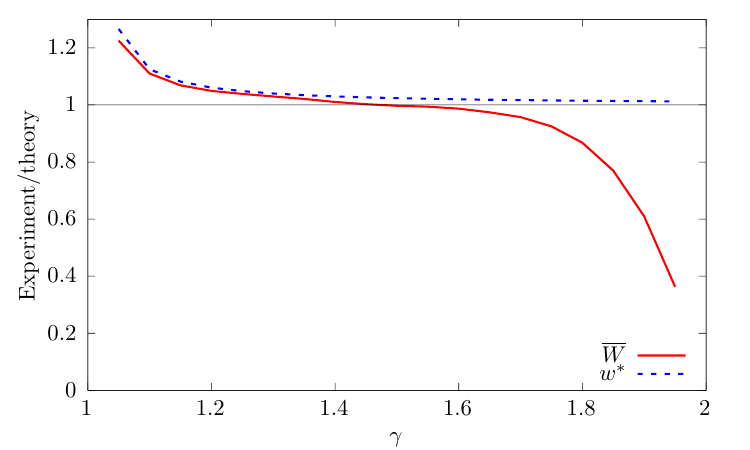}
  \caption{Plot of ratios of values of $w^\ast$ (median of the largest
    element) and $\overline W$ (median of sum of $N$ samples) obtained
    from simulation, divided by their theoretical estimates,
    equations~(\ref{eq:w_ast}) and~(\ref{eq:bar_W_estimate})
    respectively, as a function of $\gamma$.  The figure shows data
    for $N=1000$, averaged over $M=10^4$ iterations. The values of
    $\overline W$ used to generate this figure span more that $58$
    decades.}
\label{fig:ratios}
\end{figure}

The prediction for $\overline W$, equation~(\ref{eq:bar_W_estimate}),
was tested numerically. Figure \ref{fig:ratios} shows the ratio of the
empirically determined values of $w^\ast$ and $\overline W$ to the
theoretical estimates, equations~(\ref{eq:w_ast})
and~(\ref{eq:bar_W_estimate}), for $N=1000$ with $M=10^4$
realisations. This verifies equation (\ref{eq:w_ast}), and shows that
the $N$-dependence of $\overline W$ is the same as that of
$w^\ast$. The values of $\overline W$ used to create figure
\ref{fig:ratios} span many decades: theoretical values of
$\overline W$ (with $N=1000$) range from $1.58\ldots\times 10^{63}$
for $\gamma=1.05$ to $2.99\ldots\times 10^4$ at $\gamma=1.95$. Given
this very wide range of values, figure \ref{fig:ratios} demonstrates
that equation (\ref{eq:bar_W_estimate}) is a useful approximation.

\begin{figure}
\includegraphics{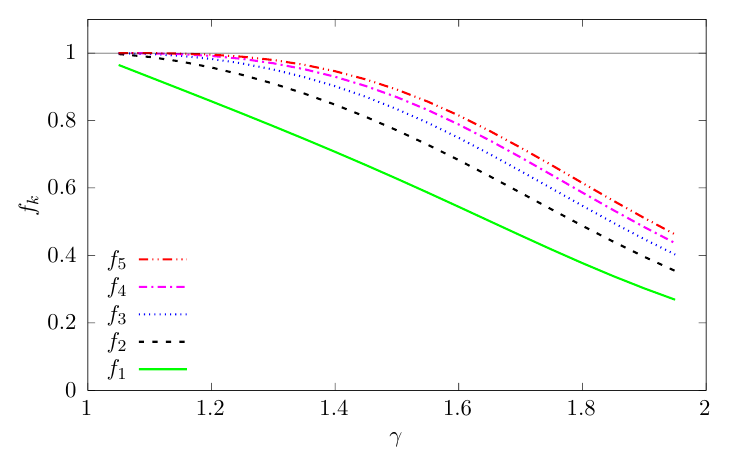}
\caption{Mean values of fraction of the sum $W$ contained in its
  largest $k$ elements (defined in equation~(\ref{eq:f_k})), as a
  function of $\gamma$. The number of elements of the sum was
  $N=1000$, and there were $M=10^4$ realisations.}
\label{fig:f_k}
\end{figure}

Figure \ref{fig:f_k} shows the expectation value of the fraction $f_k$ of the contribution to $W$
from the largest $k$ samples, as defined by (\ref{eq:f_k}). 
Figure \ref{fig:f_k} shows the expectation values $\langle f_k\rangle$ (again using $N=1000$ elements 
in the sum, and $M=10^4$ realisations) 
as a function of $\gamma$. This verifies that, in a typical realisation, most of the 
contribution to $W$ comes from a small number of the largest $w_i$. The fractional 
contribution approaches unity, in accord with the large jump 
principle \cite{Vez+19}, as $\gamma\to 1$

We remark that there is a further level of self-similarity in our
power law model, which is concerned with varying the exponent
$\gamma$. Because equation~(\ref{eq:p(w)}) implies that
$y=(\gamma-1) \ln w$ has a PDF proportional to $\exp(y)$, the
ensembles for different values of $\gamma$ are equivalent, if we
replace $w$ by $(\gamma-1)\ln w$.

Our most general conclusion from this calculation follows from
equation~(\ref{eq:bar_W_estimate}).  When extended to $d$ dimensions,
we infer that
\begin{equation}
\label{eq:bar_W_est3}
\overline W\sim R^{d/(\gamma -1)}
\end{equation}
so that the apparent dimension $D_\text{eff}$ which characterises the
scale-invariance is given by equation~(\ref{eq:Deff1}).  Note that
$D_\text{eff}>d$. We say that this scale-invariance is
\emph{ultradimensional}.  It is clearly distinguished from the
self-similarity of fractal sets, where the dimension $D$ satisfies
$D<d$.

\section{Diffusion model}
\label{sec:diffusion}

\subsection{Defining the model}
\label{sec:diff_model}

We now consider a physically motivated example of a distribution of
hotspots: the equilibrium probability density for diffusion in a
random potential, $V(\mathbf{x})$.  This example will exhibit
an approximate, rather than exact, scale-invariance.  Motion of a
particle is determined by a stochastic differential equation:
\begin{equation}
\label{eq:motion}
\delta x_i=-\mu \frac{\partial V}{\partial x_i}\delta t+
\sqrt{2\mathcal{D}}\delta \eta_i(t),
\end{equation}
where $\mu$ is the mobility and $\delta \eta_i(t)$ are white noise
signals, independent at each timestep, satisfying
$\langle \delta \eta_i\rangle=0$ and
$\langle \delta \eta_i \delta \eta_j\rangle=\delta_{ij}\delta t$.  In
the following we set $\mu=1$ throughout.  When $V=\mathrm{const}$,
the motion is simple diffusion with the diffusion coefficient
$\mathcal{D}$.  The equilibrium probability density function for the 
stochastic process (\ref{eq:motion}) is
\begin{equation}
\label{eq:P(x)}
P(\mathbf{x})=\frac{1}{\mathcal{Z}}\exp\left[-V(\mathbf{x})/\mathcal{D}\right], 
\end{equation}
where $\mathcal{Z}$ is the partition function. We shall assume that
motion is confined to a finite but very large region (which we take to
be a square with the side $\mathcal{R}$).  When $\mathcal{D}$ is small, this
density is very strongly concentrated in minima of the potential
$V(\mathbf{x})$.  Our aim will be to characterise the function $F$,
defined by equation (\ref{eq:ln_bar_w}), for this model.

\begin{figure}
  \centering
  \splot[0.45]{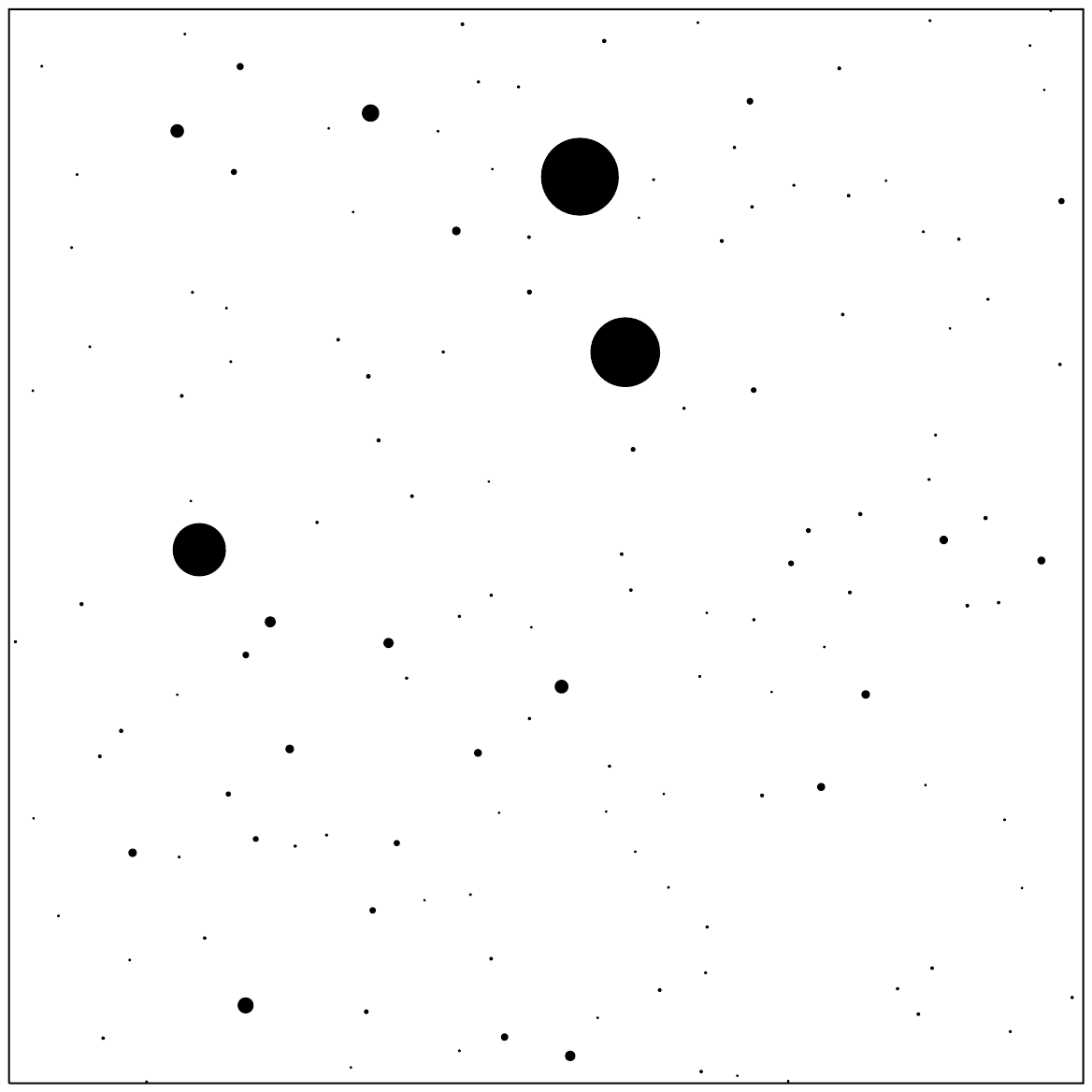}{a}
  \splot[0.45]{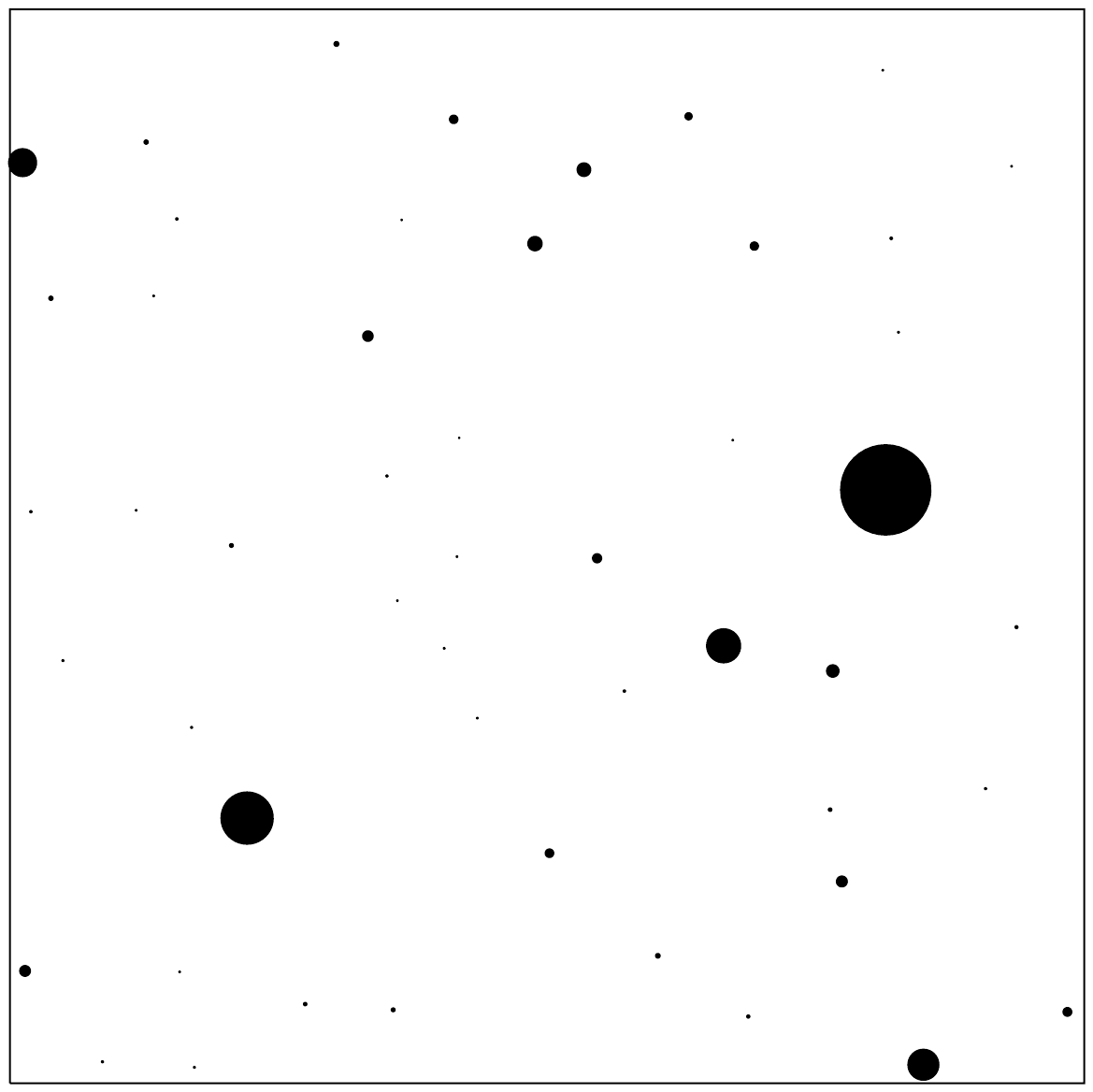}{b}
  \splot[0.45]{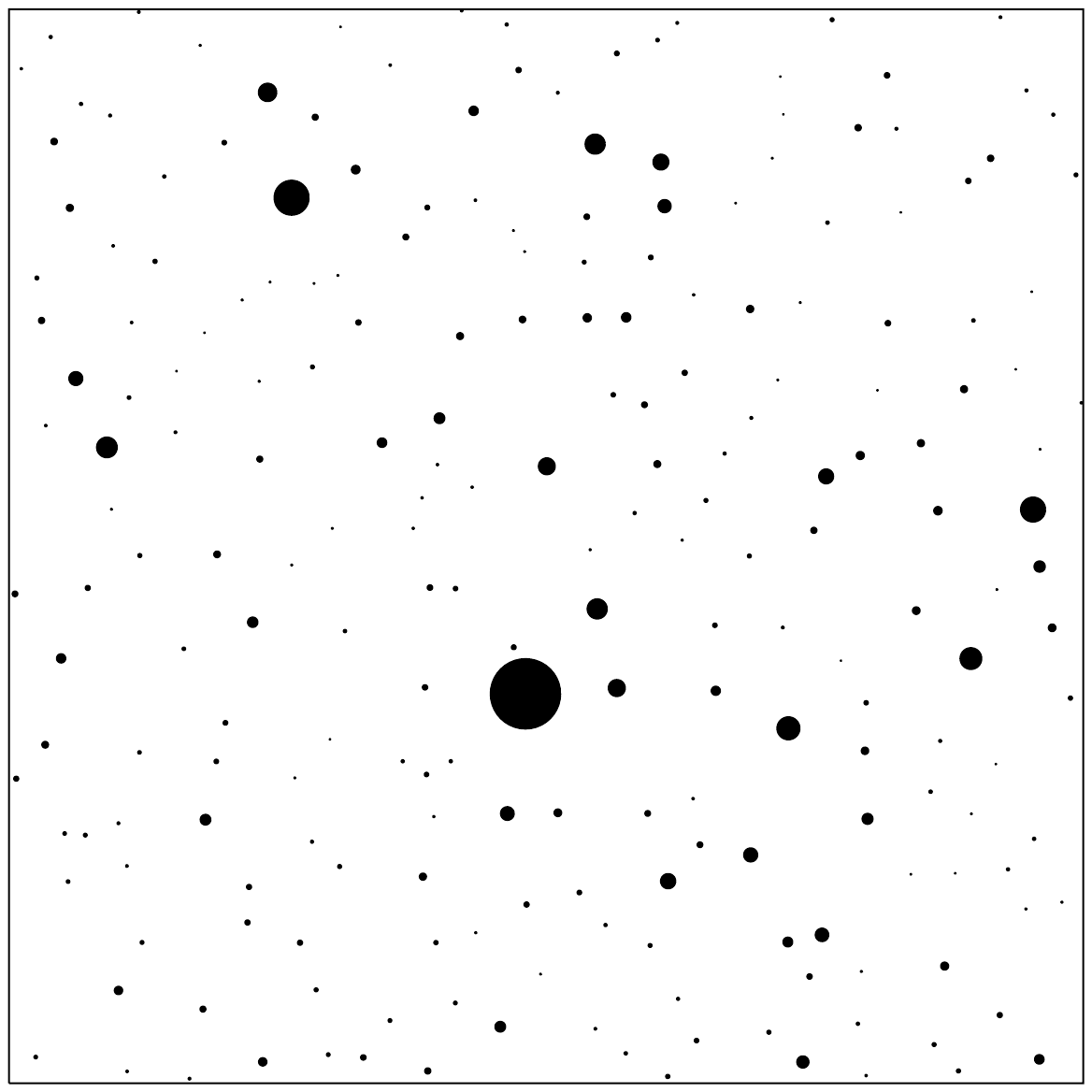}{c}
  \splot[0.45]{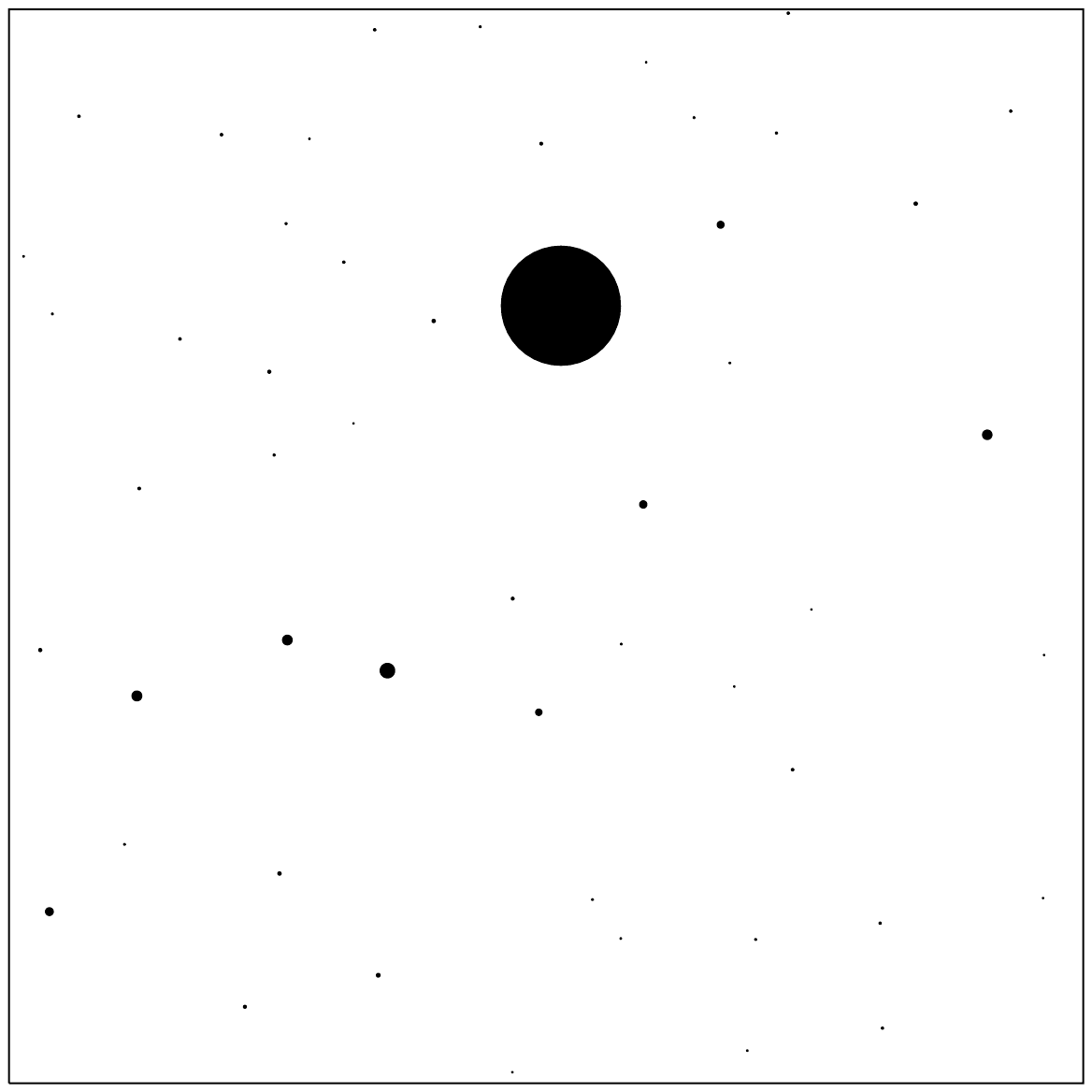}{d}
  \caption{Equilibrium probability density for diffusion in a
    two-dimensional random potential, as specified by
    (\ref{eq:Vmodel}) and (\ref{eq:c}) with a Gaussian correlation
    function,
    $\langle V(\mathbf{x})V(\mathbf{x}')\rangle=
    \exp[-(\mathbf{x}-\mathbf{x}')^2/2]$, so that $c=1$ in
    (\ref{eq:c}).  The presentation is the same as in
    figure~\ref{fig:hotspots}: hotspots are represented by circles by
    the areas proportional to their weights, with $1\%$ of the image
    covered. Panel (a): $\mathcal{D}=0.25$, $R=50$. Panel (b):
    $\mathcal{D}=0.25$, $R=25$. Panel (c): $\mathcal{D}=0.4$, $R=50$.
    Panel (d): $\mathcal{D}=0.4$, $R=25$.}
  \label{fig:diffusion}
\end{figure}

Consider the equilibrium measure when the potential $V(\mathbf{x})$ is
itself a smoothly varying random function, with a Gaussian PDF, and
statistics which are homogeneous and isotropic. We shall assume that
$V(x,y)$ has the following statistical properties:
\begin{equation}
\label{eq:Vmodel}
\langle V\rangle=0,\quad
\langle V^2\rangle=1,\quad
\langle V_x^2\rangle=\langle V_y^2\rangle=1,
\end{equation}
where $V_x=\partial V/\partial x$, etc.  These requirements can be
satisfied by re-scaling the coordinates and the potential.  Also
define $c$ by writing
\begin{equation}
\label{eq:c}
c=\langle V_{xy}^2\rangle.
\end{equation}
This parameter satisfies $c\ge 1/2$, with the lower limit realised if
the spectral function $S(k)$ of $V(\mathbf{x})$ (the modulus squared
of the Fourier transform of its autocorrelation) has a ring spectrum,
$S(k)\propto \delta (k-k_0)$. If the correlation function of $V$ is a
Gaussian, then $c=1$.

In general, the value of $\mathcal{Z}$ depends upon the realisation of
the potential $V(\mathbf{x})$, but if the scale size $\mathcal{R}$ of
the region is sufficiently large, we can use ergodicity and
approximate $\mathcal{Z}$ by its expectation value,
$\langle \mathcal{Z}\rangle$.  The partition function is then
approximated by
\begin{equation}
  \label{eq:Z}
  \begin{aligned}
    \langle \mathcal{Z}\rangle &=\mathcal{R}^2\left\langle
      \exp\left[-\frac{V}{\mathcal{D}}\right]\right\rangle \\
    &=\frac{\mathcal{R}^2}{\sqrt{2\pi }}\int_{-\infty}^\infty
    \diff V\,\exp\left[-\frac{V^2}{2}-\frac{V}{\mathcal{D}}\right] \\
    &=\exp\left[\frac{1}{2\mathcal{D}^2}\right]\mathcal{R}^2.
\end{aligned}
\end{equation}
When $\mathcal{D}$ is sufficiently small that the measure (\ref{eq:P(x)}) is
concentrated at the minima of $V(\mathbf{x})$, the weight of a 
hotspot is approximated by
\begin{equation}
\label{eq:w}
w=\frac{1}{\mathcal{Z}}\int \diff x\int \diff y\ \exp\left[-V(x,y)/{\mathcal{D}}\right] 
\sim\frac{2\pi \mathcal{D}}{\mathcal{Z}} \Delta^{-1/2} \exp[-V^\ast/{\mathcal{D}}],
\end{equation}
where $V^\ast$ is the height of the minimum, and
$\Delta=V_{xx}V_{yy}-V_{xy}^2$ is the determinant of the Hessian
matrix at the minimum.

Figure~\ref{fig:diffusion} illustrates the distribution of the weights
of the hotspots of this diffusion model, using the same presentation
as figure~\ref{fig:hotspots} (hotspots are represented by a filled
circle with the area proportional to its weight,
equation~(\ref{eq:w}), and the total area of circles is normalised to
$1\%$).  We used two different diffusion coefficients $\mathcal{D}$
and lengthscales $R$.  The distributions are qualitatively similar to 
those of the simplified model, shown in figure~\ref{fig:hotspots}.

When $\mathcal{D}$ is small, the weights of the hotspots have a very
broad distribution. The expectation value $\langle w\rangle$ is
dominated by extremely rare events, which are unlikely to be realised,
and it is more useful to estimate the median $\overline W$ of the
total weight inside a region of area $R^2$. The growth of
$\overline W$ as a function of $R$ is characterised by calculating the
function $F$ defined by equation~(\ref{eq:ln_bar_w}):
$\ln \overline W=F(\ln R)$.  It will be argued that, for this model,
the large-jump principle \cite{Vez+19} is applicable, so that
$\overline W$ is well approximated by the median of its largest
contributor, denoted by $w^\ast$.

We shall consider the following scenario. The potential
$V(\mathbf{x})$ is evaluated, and the weights (\ref{eq:w}) calculated,
in a region of size $\mathcal{R}$. While $\mathcal{R}$ is assumed to
be large, we assume that $\mathcal{D}$ is sufficiently small that
$\mathcal{Z}\ll \langle \mathcal{Z}\rangle$, so that the largest
weight is $\hat w\approx 1$. This implies that when we estimate
$\overline W(R)$, our estimate should satisfy
$\overline W(\mathcal{R})=1$. We assume that the density of minima of
$V(\mathbf{x})$ is $\rho$.  According to equation (\ref{eq:w}), a
large value of $w$ is associated with a minimum of the potential $V$,
which has an approximate depth
$V\approx -\mathcal{D}[\ln w+\ln \mathcal{Z}]$, and we find it
convenient to use a variable
\begin{equation}
\label{eq:Vtilde}
\widetilde V\equiv -\mathcal{D}\left[\ln w+\ln \mathcal{Z} \right]
\end{equation}
instead of $w$, because the distribution of weights has a narrow
support when expressed in terms of $\widetilde V$. The largest values
of $w$ are observed very rarely, so we shall characterise the density
of hotspots with very large values of $w$ as follows: the probability
$P(\widetilde V_0)$ that $\widetilde V$ is less than $\widetilde V_0$
is written in the form
\begin{equation}
\label{eq:P(tildeV)}
P(\widetilde V)=\exp\left[-J(\widetilde V)\right].
\end{equation}
(In order to unambiguously normalise this distribution we regard any
minimum of $V(\mathbf{x})$ as being a hotspot).  The function $J(V)$
corresponds to a `rate function' or `entropy function' of large
deviation theory \cite{Tou09}. We can then estimate the median of the
smallest value of $\widetilde V$, denoted by $\widetilde V^\ast$, by
writing $1/2=[1-P(\widetilde V^\ast)]^{\rho R^2}$, where $\rho $ is
the density of minima. This yields:
\begin{equation}
\label{eq:J}
J(\widetilde V^\ast)=2\ln R+\ln \rho - \ln \ln 2.
\end{equation}
If the inverse function of $J$ is $K$ (that is $K(J(V))=V$), then the required 
relation between $w^\ast=\overline W$ and $\ln R$ is 
\begin{equation}
\label{eq:bar_W_4}
\ln \overline W=-\ln \mathcal{Z}-\frac{1}{\mathcal{D}}K(2\ln R+\ln
\rho -\ln \ln 2)\equiv F(\ln R). 
\end{equation}
In order to use this expression to determine the function $F(\ln R)$
which appears in equation (\ref{eq:ln_bar_w}), we must determine the
large-deviation rate function $J(V)$ which was introduced in
equation~(\ref{eq:P(tildeV)}).

Note that, according to equations~(\ref{eq:w}) and~(\ref{eq:Vtilde}), 
\begin{equation}
\label{eq:V_est1}
\widetilde V=V^\ast  +\mathcal{D}\ln \left(\frac{\sqrt{\Delta}}{2\pi
    \mathcal{D}}\right), 
\end{equation}
so that, in the limit as $\mathcal{D}\to 0$, $\widetilde V\to V$, and
it is sufficient for our purposes to determine the PDF of the heights
of minima of the function $V(x,y)$.  We can, therefore, use the
cumulative probability of the heights of local minima as the function
$P$ in equation (\ref{eq:P(tildeV)}).

Equations~(\ref{eq:gamma_eff}) and~(\ref{eq:bar_W_4}) imply that
\begin{equation}
\label{eq:gamma_eff1}
\gamma_\text{eff}\sim 1-\mathcal{D}J'(\widetilde V^\ast),
\end{equation}
so that $\gamma_\text{eff}\sim 1$ when $\mathcal{D}\to 0$. This
observation justifies the claim that $\hat W\sim w^\ast$.

\subsection{Distribution of weights}
\label{sec:distribution_weights}

We now turn to evaluating the distribution of heights of minima. The
two-dimensional case is quite technical, so we shall start by
discussing the estimate of $\overline W(R)$ in one dimension.

Here we require the density of local minima, $\rho$, and the
probability $P(V)$ that the height of a local minimum is less than
$V$. These are readily obtained using the approach developed by Rice
\cite{Ric45}, following pioneering work by Kac~\cite{Kac43}.  The
density of minima is
\begin{equation}
\label{eq:rho}
\rho=\int_{-\infty}^\infty \diff V\int_0^\infty \diff V''\, V''P(V,0,V''),
\end{equation}
where $P(V,V',V'')$ is the joint PDF of $V$ and its first two
derivatives, evaluated at the same point.  We consider the case where
$V(x)$ is Gaussian, with correlation function
\begin{equation}
\label{eq:VV'}
\langle V(x)V(x')\rangle=\exp[-(x-x')^2/2].
\end{equation}
We find the following non-zero statistics of the potential and its
derivatives at a given point:
$\langle V^2\rangle=\langle V'^2\rangle=1$, $\langle V''^2\rangle=3$,
$\langle VV''\rangle=-1$. Using the standard formula for multivariate
Gaussian distribution, we find
\begin{equation}
\label{eq:PVV'V''}
P(V,V',V'')=\frac{1}{4\pi^{3/2}}\exp\left[-\frac{1}{4}\left(3V^2+2VV''+V''^2\right)\right]\exp(-V'^2/2), 
\end{equation}
and hence the density of minima is
\begin{equation}
\label{eq:rho-sol}
\rho=\frac{\sqrt{3}}{2\pi}.
\end{equation}

\begin{figure}
  \centering
  \parbox{0.45\textwidth}{\centering
    \includegraphics[width=0.45\textwidth]{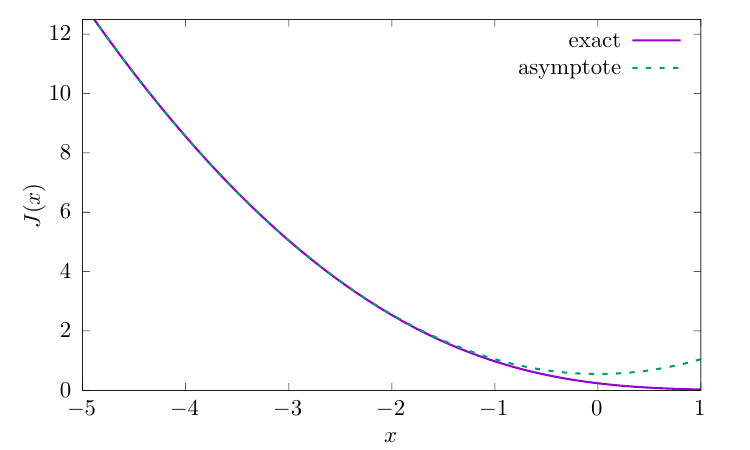}\\(a)}
  \parbox{0.45\textwidth}{\centering
    \includegraphics[width=0.45\textwidth]{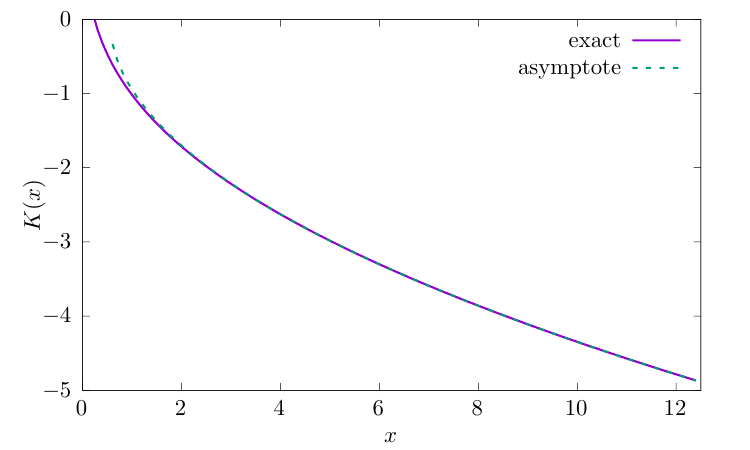}\\
    (b)}
\caption{(a) The large deviation function for the distribution of
  minima, $J(V)$ (defined by equation~(\ref{eq:P(tildeV)})), for a
  one-dimensional random potential with Gaussian correlation function,
  $\langle V(x)V(x')\rangle=\exp[-(x-x')^2/2]$.  The large deviation
  rate function $J(V)$ derived from the cumulative distribution of
  minima $P(V)$, as defined by equation~(\ref{eq:P(V)1}), is shown as
  a solid line.  Its asymptotic approximation, (\ref{eq:JK}), is shown
  as a dashed line.  (b) The inverse function, $K(J)$: exact
  shown as solid line, asymptote~(\ref{eq:JK}) is dashed line.}
\label{fig:JK1D}
\end{figure}

The PDF of the heights of minima is
\begin{equation}
  \label{eq:p(V)}
  \begin{aligned}
        p(V)&=\frac{1}{\rho}\int_0^\infty \diff V''\, V''P(V,0,V'')\\
    &=\frac{1}{\sqrt{3\pi}}\exp(-3V^2/4)-\frac{1}{2\sqrt{3}}V\exp(-V^2/2)\erfc
    \left(\frac{V}{2}\right)
  \end{aligned}
\end{equation}
and the cumulative probability for the minimum being at a level 
less that $V$ is 
\begin{equation}
\label{eq:P(V)1}
P(V)=\int_{-\infty}^V \diff v\, p(v)
=\frac{1}{2}\left[1+\erf\left(\frac{\sqrt{3}V}{2}\right)\right]
+\frac{1}{2\sqrt{3}}\exp(-V^2/2)\erfc \left(\frac{V}{2}\right)
\end{equation}
which also allows us to obtain $J(V)=-\ln[P(V)]$ explicitly. 
The asymptote for $J(V)$ as
$V\to -\infty$, and the corresponding asymptote for its inverse
function $K(J)$ are:
\begin{equation}
\label{eq:JK}
J(V)\sim \frac{V^2}{2}+\frac{\ln 3}{2},\quad
K(J)\sim \sqrt{2J-\ln 3}.
\end{equation}
The functions $J(V)$ and $K(J)$ for the one-dimensional model with Gaussian 
correlation function ($c=1$) are plotted in figure \ref{fig:JK1D}.

\begin{figure}
  \centering
  \parbox{0.45\textwidth}{\centering
    \includegraphics[width=0.45\textwidth]{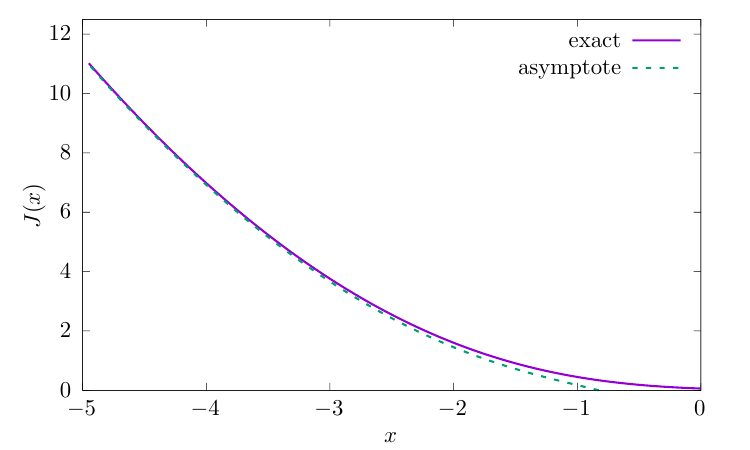}\\(a)}
  \parbox{0.45\textwidth}{\centering
    \includegraphics[width=0.45\textwidth]{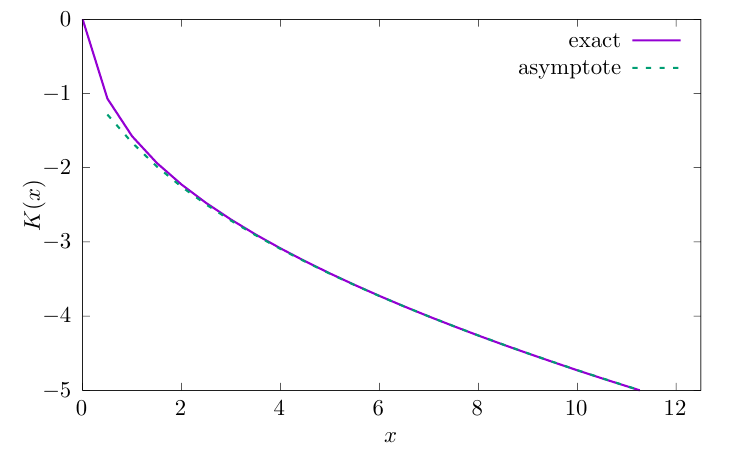}\\
    (b)}
  \caption{(a) The large deviation function for the distribution of
    minima, $J(V)$ (defined by equaiton~(\ref{eq:P(tildeV)})), for a
    two-dimensional random potential with Gaussian correlation
    function,
    $\langle V(\mathbf{x})V(\mathbf{x}')\rangle
    =\exp[-(\mathbf{x}-\mathbf{x}')^2/2]$.  The large deviation rate
    function $J(V)$ derived from the cumulative distribution of minima
    $P(V)$, obtained by numerical integration of
    equation~(\ref{eq:p(V)2D}), is shown as a solid line. Its
    asymptotic approximation, (\ref{eqJK2D}), is shown as a dashed
    line.  (b) The inverse function, $K(J)$: exact shown as solid
    line, asymptote~(\ref{eqJK2D}) is dashed line.}
\label{fig:JK2D}
\end{figure}

In the two-dimensional case the calculation of the distribution is
more difficult, but the result is already known: for the case where
the correlation function is a Gaussian, the PDF of the distribution of
minima is \cite{Wil+92} (see also erratum, \cite{Wil+94}):
\begin{multline}
\label{eq:p(V)2D}
p(V) =\frac{\sqrt{3}}{2\pi}\bigg[
\sqrt{\pi}\exp\left(-\frac{3}{4}V^2\right)\erfc \left(\frac{V}{2}\right)  -V\exp\left(-V^2\right)\\
  +\sqrt{\frac{\pi}{2}}(V^2-1)\exp\left(-\frac{V^2}{2}\right)\erfc \left(\frac{V}{\sqrt{2}}\right)
  \bigg]
\end{multline}
and the density of minima in two dimensions is \cite{Wil+92}
\begin{equation}
\label{eq:rho2D}
\rho=\frac{1}{2\pi \sqrt{3}}.
\end{equation}
The corresponding cumulative distribution cannot be expressed in terms
of familiar special functions, so we obtained $P(V)$ by numerical
integration. The asymptote can, however, be determined analytically:
\begin{equation}
  \label{eqJK2D}
  \begin{aligned}
J(V)&\sim
\frac{V^2}{2}-\ln\left(\frac{V^2+1}{|V|}\right)+\frac{1}{2}\ln\left(\frac{2\pi}{3}\right)\\ 
K(J)&\sim
\sqrt{2J+\ln\left({2J+1}{\sqrt{2J}}\right)+\frac{1}{2}\ln\left(\frac{3}{2\pi}\right)}.
\end{aligned}
\end{equation}
Figure \ref{fig:JK2D} shows the corresponding function $J(V)$ and its
inverse, compared with (\ref{eqJK2D}).

We investigated the statistics of hotspots for the model of diffusion
in Gauss random potential (with Gaussian correlation function), by
evaluating the function $F(x)$ defined by
equation~(\ref{eq:ln_bar_w}), for different values of the diffusion
coefficient $\mathcal{D}$.  Because we were able to perform numerical
simulations over a wider range of $R$-values in the one-dimensional
case, we present results for both the one- and two-dimensional models.

\begin{figure}
  \centering
\includegraphics{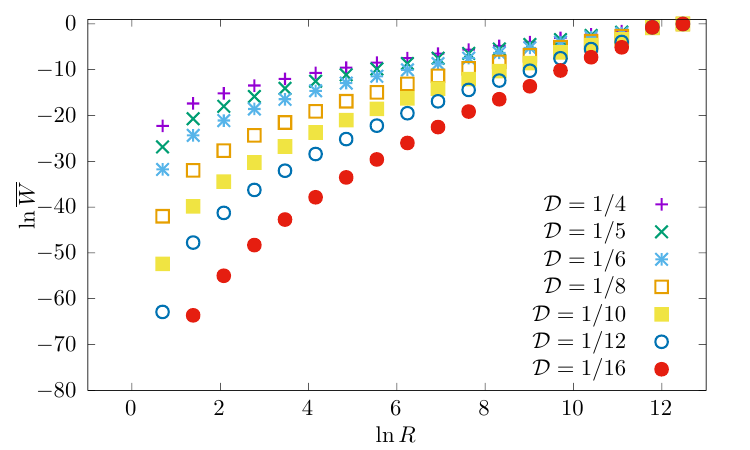}
\caption{Numerical investigation of the function $F(x)$ defined in
  equation~(\ref{eq:ln_bar_w}) for the one-dimensional model with
  Gaussian correlation function: $\ln \overline W$ is evaluated as a
  function of $\ln R$, for a range of different values of
  $\mathcal{D}$.}
\label{fig:lnW1D}
\end{figure}

\begin{figure}
  \centering
\includegraphics{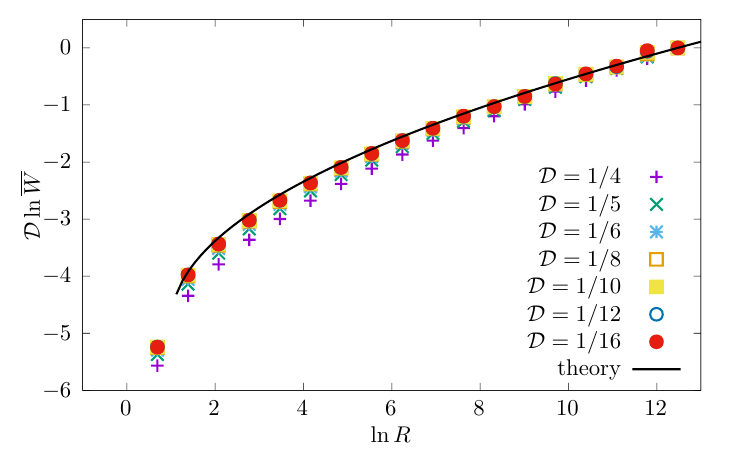}
\caption{The data in figure \ref{fig:lnW1D} collapse onto the function
  $K(x)$ plotted in figure \ref{fig:JK1D}(b), in accord with
  equation~(\ref{eq:bar_W(R)}).}
\label{fig:lnW1Dcollapse}
\end{figure}

The results for the one-dimensional case are summarised in figures
\ref{fig:lnW1D} and \ref{fig:lnW1Dcollapse}. For each value of $\mathcal{D}$,
we generated $M=20$ realisations of the random potential $V(x)$ on an
interval of length $L=100\times 2^{18}$, by smoothing white noise
using a Gaussian kernel.  The partition function $\mathcal{Z}$ was
calculated for each realisation, and the local minima $x_i$ of the
potential were identified, together with the values of $V(x_i)$ and
$V''(x_i)$.  For each realisation, we divided the interval into
sub-intervals, halving the length each time, for $18$ generations. At
generation $k=1$,\ldots,$18$, for each of the $M\times 2^{k-1}$
sub-intervals of length $R=L/2^{k-1}$, we sum the weights $w$ to
determine the total weight $W$ of each sub-interval. We then
determined the median values, $\overline W$, of these
$M\times 2^{k-1}$ weights.  In figure \ref{fig:lnW1D} we plot the
resulting $18$ values of $\ln \overline W$ as a function of $\ln R$,
for several different values of the diffusion coefficient
$\mathcal{D}$.

Because the values of $\mathcal{R}$ and $\mathcal{D}$ were chosen so
that the largest weights $w_i$ were of order unity,
equation~(\ref{eq:bar_W_4}) simplifies to
\begin{equation}
\label{eq:bar_W(R)}
\ln \overline W(R)=
\frac{1}{\mathcal{D}}K\left(\ln \mathcal{R}+\ln \rho -\ln \ln 2\right)
-\frac{1}{\mathcal{D}}K\left(\ln R+\ln \rho -\ln \ln 2\right),
\end{equation}
where $\rho$ is the density of minima ((\ref{eq:rho-sol}) in one
dimension, (\ref{eq:rho2D}) in two dimensions).  Figure~\ref{fig:lnW1Dcollapse}
verifies this expression by showing a collapse of the data in
figure~\ref{fig:lnW1D} onto the inverse function of the
large-deviation entropy, $K(x)$ in the one-dimensional case.

\begin{figure}
\includegraphics[width=12cm]{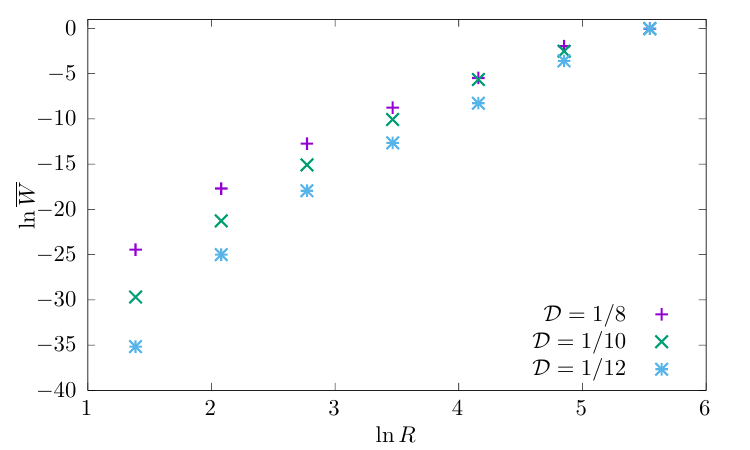}
\caption{ Numerical investigation of the function $F(x)$ defined in
  equation~(\ref{eq:ln_bar_w}) for the two-dimensional model with
  Gaussian correlation function: $\ln \overline W$ is evaluated as a
  function of $\ln R$, for a range of different values of
  $\mathcal{D}$.}
\label{fig:lnW2D}
\end{figure}

\begin{figure}
\includegraphics[width=12cm]{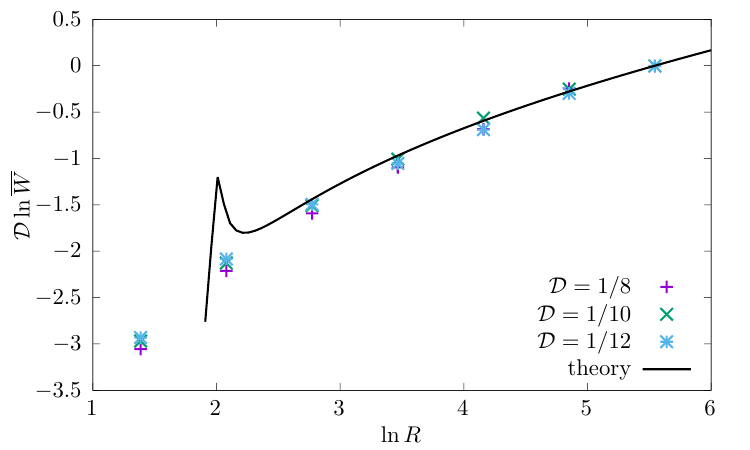}
\caption{The data in figure \ref{fig:lnW2D} collapse onto the function
  $K(x)$ plotted in figure \ref{fig:JK2D}(b), in accord with
  equation (\ref{eq:bar_W(R)}).  Note that the assumptions behind
  asymptotics~(\ref{eq:bar_W(R)}) break at $\ln R<2.5$, but the data
  still collapse even at smaller $R$.}
\label{fig:lnW2D-collapse}
\end{figure}

We generated $M=4$ realisations of $V(x,y)$ on a square of size ${\cal R}=256$, 
with toroidal boundary conditions, by convoluting a discrete representation of white noise
with a Gaussian kernel.
Figure~\ref{fig:lnW2D} displays plots of $\ln \overline W$ as a
function of $\ln R$ for the two-dimensional Gaussian potential, with
different values of the diffusion coefficient $\mathcal{D}$, for $R=2,4,\ldots,256$.  Figure
\ref{fig:lnW2D-collapse} illustrates the collapse of these data onto a
plot of $K(x)$, the inverse of the large-deviation rate function
$J(x)$. The range of values of $R$ is much smaller than that shown in
figures \ref{fig:lnW1D} and \ref{fig:lnW1Dcollapse}, because the
two-dimensional simulations are more numerically demanding.

\section{Concluding remarks}
\label{sec:conclusion}

Images which show the distribution of `hotspots', where a field has an
unusually high intensity, appear to have a family resemblance, which
may not be strongly dependent upon the size of the sample region.

The distribution of weights of hotspots was characterised by
considering the median $\overline W$ of the total weight in a region
of size $R$, and defining a function $F(x)$ by writing
$\ln \overline W=F(\ln R)$ (equation~(\ref{eq:ln_bar_w})). The
derivative of $F$ is an effective dimension, $D_\text{eff}=F'(\ln R)$.

We investigated two models for hotspot distributions.  Firstly, we
considered a one-parameter family of weight distributions, defined 
by~(\ref{eq:p(w)}), which were contrived to be
scale-invariant. The scale-invariance of these models is characterised
by an effective dimension $D_\text{eff}=d/(\gamma-1)$, where
$\gamma \in (1,2)$ is the parameter in the definition of the model.
Because this dimension is greater than that of the embedding space,
the scale invariance is distinct from the self-similarity which
characterises fractal sets.  Examples of realisations of this model
are shown in figure \ref{fig:hotspots}.  While it is a mathematical
fact that the individual images are drawn from the same ensemble, the
realisations do look very different from each other.

We also considered a physically motivated example, namely the
equilibrium distribution for diffusion in a random potential. Here the
realisations of the hotspot distribution, illustrated in
figure~\ref{fig:diffusion}, are qualitatively similar to those of the
simplified model. We were able to determine the function $F(x)$ for
this model. Because it is not a linear function, this system does not
exhibit strict scale invariance.

\bmhead{Acknowledgments}

This work was initiated while MW was a guest of the 
Chan-Zuckerberg Biohub. 

\bmhead{Data availability}

No externally sourced data was processed. The numerical codes used for the simulations 
are available from the corresponding author. 


\begin{thebibliography}{9}
\ifx \bisbn   \undefined \def \bisbn  #1{ISBN #1}\fi
\ifx \binits  \undefined \def \binits#1{#1}\fi
\ifx \bauthor  \undefined \def \bauthor#1{#1}\fi
\ifx \batitle  \undefined \def \batitle#1{#1}\fi
\ifx \bjtitle  \undefined \def \bjtitle#1{#1}\fi
\ifx \bvolume  \undefined \def \bvolume#1{\textbf{#1}}\fi
\ifx \byear  \undefined \def \byear#1{#1}\fi
\ifx \bissue  \undefined \def \bissue#1{#1}\fi
\ifx \bfpage  \undefined \def \bfpage#1{#1}\fi
\ifx \blpage  \undefined \def \blpage #1{#1}\fi
\ifx \burl  \undefined \def \burl#1{\textsf{#1}}\fi
\ifx \doiurl  \undefined \def \doiurl#1{\url{https://doi.org/#1}}\fi
\ifx \betal  \undefined \def \betal{\textit{et al.}}\fi
\ifx \binstitute  \undefined \def \binstitute#1{#1}\fi
\ifx \binstitutionaled  \undefined \def \binstitutionaled#1{#1}\fi
\ifx \bctitle  \undefined \def \bctitle#1{#1}\fi
\ifx \beditor  \undefined \def \beditor#1{#1}\fi
\ifx \bpublisher  \undefined \def \bpublisher#1{#1}\fi
\ifx \bbtitle  \undefined \def \bbtitle#1{#1}\fi
\ifx \bedition  \undefined \def \bedition#1{#1}\fi
\ifx \bseriesno  \undefined \def \bseriesno#1{#1}\fi
\ifx \blocation  \undefined \def \blocation#1{#1}\fi
\ifx \bsertitle  \undefined \def \bsertitle#1{#1}\fi
\ifx \bsnm \undefined \def \bsnm#1{#1}\fi
\ifx \bsuffix \undefined \def \bsuffix#1{#1}\fi
\ifx \bparticle \undefined \def \bparticle#1{#1}\fi
\ifx \barticle \undefined \def \barticle#1{#1}\fi
\bibcommenthead
\ifx \bconfdate \undefined \def \bconfdate #1{#1}\fi
\ifx \botherref \undefined \def \botherref #1{#1}\fi
\ifx \url \undefined \def \url#1{\textsf{#1}}\fi
\ifx \bchapter \undefined \def \bchapter#1{#1}\fi
\ifx \bbook \undefined \def \bbook#1{#1}\fi
\ifx \bcomment \undefined \def \bcomment#1{#1}\fi
\ifx \oauthor \undefined \def \oauthor#1{#1}\fi
\ifx \citeauthoryear \undefined \def \citeauthoryear#1{#1}\fi
\ifx \endbibitem  \undefined \def \endbibitem {}\fi
\ifx \bconflocation  \undefined \def \bconflocation#1{#1}\fi
\ifx \arxivurl  \undefined \def \arxivurl#1{\textsf{#1}}\fi
\csname PreBibitemsHook\endcsname

\bibitem[\protect\citeauthoryear{Mandelbrot}{1983}]{Man83}
\begin{bbook}
\bauthor{\bsnm{Mandelbrot}, \binits{B.B.}}:
\bbtitle{The Fractal Geometry of Nature},
\bedition{3}rd edn.
\bpublisher{W. H. Freeman and Comp.},
\blocation{New York}
(\byear{1983})
\end{bbook}
\endbibitem

\bibitem[\protect\citeauthoryear{Falconer}{1990}]{Fal90}
\begin{bbook}
\bauthor{\bsnm{Falconer}, \binits{K.}}:
\bbtitle{Fractal Geometry---Mathematical Foundations and Applications.},
\bpublisher{Wiley},
\blocation{New York}
(\byear{1990})
\end{bbook}
\endbibitem

\bibitem[\protect\citeauthoryear{Halsey et~al.}{1986}]{Hal+86}
\begin{barticle}
\bauthor{\bsnm{Halsey}, \binits{T.C.}},
\bauthor{\bsnm{Jensen}, \binits{M.H.}},
\bauthor{\bsnm{Kadanoff}, \binits{L.P.}},
\bauthor{\bsnm{Procaccia}, \binits{I.}},
\bauthor{\bsnm{Shraiman}, \binits{B.I.}}:
\batitle{Fractal measures and their singularities: The characterization of
  strange sets}.
\bjtitle{Phys. Rev. A}
\bvolume{33},
\bfpage{1141}--\blpage{1151}
(\byear{1986})
\doiurl{10.1103/PhysRevA.33.1141}
\end{barticle}
\endbibitem

\bibitem[\protect\citeauthoryear{Salat et~al.}{2017}]{Sal+17}
\begin{barticle}
\bauthor{\bsnm{Salat}, \binits{H.}},
\bauthor{\bsnm{Murcio}, \binits{R.}},
\bauthor{\bsnm{Arcaute}, \binits{E.}}:
\batitle{Multifractal methodology}.
\bjtitle{Physica A: Statistical Mechanics and its Applications}
\bvolume{473},
\bfpage{467}--\blpage{487}
(\byear{2017})
\doiurl{10.1016/j.physa.2017.01.041}
\end{barticle}
\endbibitem

\bibitem[\protect\citeauthoryear{Vezzani et~al.}{2019}]{Vez+19}
\begin{barticle}
\bauthor{\bsnm{Vezzani}, \binits{A.}},
\bauthor{\bsnm{Barkai}, \binits{E.}},
\bauthor{\bsnm{Burioni}, \binits{R.}}:
\batitle{Single-big-jump principle in physical modeling}.
\bjtitle{Phys. Rev. E}
\bvolume{100},
\bfpage{012108}
(\byear{2019})
\doiurl{10.1103/PhysRevE.100.012108}
\end{barticle}
\endbibitem

\bibitem[\protect\citeauthoryear{Touchette}{2009}]{Tou09}
\begin{barticle}
\bauthor{\bsnm{Touchette}, \binits{H.}}:
\batitle{The large deviation approach to statistical mechanics}.
\bjtitle{Physics Reports}
\bvolume{478}(\bissue{1}),
\bfpage{1}--\blpage{69}
(\byear{2009})
\doiurl{10.1016/j.physrep.2009.05.002}
\end{barticle}
\endbibitem

\bibitem[\protect\citeauthoryear{Rice}{1945}]{Ric45}
\begin{barticle}
\bauthor{\bsnm{Rice}, \binits{S.O.}}:
\batitle{Mathematical analysis of random noise}.
\bjtitle{Bell System Technical Journal}
\bvolume{24}(\bissue{1}),
\bfpage{46}--\blpage{156}
(\byear{1945})
\doiurl{10.1002/j.1538-7305.1945.tb00453.x}
\end{barticle}
\endbibitem

\bibitem[\protect\citeauthoryear{Kac}{1943}]{Kac43}
\begin{barticle}
\bauthor{\bsnm{Kac}, \binits{M.}}:
\batitle{On the average number of real roots of a random algebraic equation}.
\bjtitle{Bulletin of the American Mathematical Society}
\bvolume{49},
\bfpage{314}--\blpage{320}
(\byear{1943})
\doiurl{10.1090/S0002-9904-1943-07912-8}
\end{barticle}
\endbibitem

\bibitem[\protect\citeauthoryear{Wilkinson et~al.}{1992}]{Wil+92}
\begin{barticle}
\bauthor{\bsnm{Wilkinson}, \binits{M.}},
\bauthor{\bsnm{Yang}, \binits{F.}},
\bauthor{\bsnm{Austin}, \binits{E.J.}},
\bauthor{\bsnm{O'Donnell}, \binits{K.P.}}:
\batitle{A statistical topographic model for exciton luminescence spectra}.
\bjtitle{Journal of Physics: Condensed Matter}
\bvolume{4}(\bissue{45}),
\bfpage{8863}
(\byear{1992})
\doiurl{10.1088/0953-8984/4/45/019}
\end{barticle}
\endbibitem

\bibitem[\protect\citeauthoryear{Wilkinson et~al.}{1994}]{Wil+94}
\begin{barticle}
\bauthor{\bsnm{Wilkinson}, \binits{M.}},
\bauthor{\bsnm{Yang}, \binits{F.}},
\bauthor{\bsnm{Austin}, \binits{E.J.}},
\bauthor{\bsnm{O'Donnell}, \binits{K.P.}}:
\batitle{Corrigendum}.
\bjtitle{Journal of Physics: Condensed Matter}
\bvolume{6}(\bissue{16}),
\bfpage{3123}
(\byear{1994})
\end{barticle}
\endbibitem


\end{thebibliography}

\end{document}